\pdfoutput=1

\documentclass[]{raa}

\usepackage{mathptmx}
\usepackage{ae,aecompl}

\usepackage{graphicx,times}	
\usepackage{amsmath}	
\usepackage{amssymb}	
\usepackage{pifont}
\usepackage{txfonts}
\usepackage{url}
\usepackage{subfigure}
\usepackage{longtable}
\usepackage{lscape}
\usepackage{multirow}
\usepackage{color}
\usepackage{textcomp}
\usepackage{natbib}
\usepackage{epstopdf}
\usepackage{pdfpages}
\usepackage[colorlinks,linkcolor=red,anchorcolor=green,citecolor=blue]{hyperref}
\usepackage{soul}
\usepackage{appendix}

\begin{document}

\title{ Detection of 16 small glitches in 9 pulsars}


   \volnopage{Vol.0 (2022) No.0, 000--000}      
   \setcounter{page}{1}          
       \author{Zu-Rong Zhou
      \inst{1,2}
   \and Jing-Bo Wang
      \inst{1,3,4}
   \and Na Wang
      \inst{1,3,4}
      \and Jian-Ping Yuan
      \inst{1,3,4} 
   \and Fei-Fei Kou
      \inst{1,3,4} 
   \and Shi-Jun Dang
      \inst{5}      
 }
  

 \institute{Xinjiang Astronomical Observatory, Chinese Academy of Sciences, Urumqi, XinJiang 830011, China \\
        \and
            University of Chinese Academy of Sciences, 19A Yuquan Road, Beijing 100049, China \\
        \and
           Xinjiang Key Laboratory of Radio Astrophysics, 150 Science1-Street, Urumqi, Xinjiang, 830011, China \\
       \and 
            Key Lab of Radio Astronomy, Chinese Academy of Sciences, Beijing 100101, China \\
       \and
           School of Physics and Electronic Science, Guizhou Normal University, Guiyang 550001, China \\
    \vs\no
   {\small Received 20xx month day; accepted 2022 June 21th}}
\email{{\it wangjingbo@xao.ac.cn; na.wang@xao.ac.cn}}

\abstract{Timing observations from the Nanshan 26-m radio telescope for nine pulsars between 2000 and 2014 have been used to search for glitches. The data span for nine pulsars ranges from 11.6 to 14.2 years. From the total of 114 yr of pulsar rotational history, 16 new glitches were identified in 9 pulsars. Glitch parameters were measured by fitting the timing residuals data. All 16 glitches have a small fractional size. Six new glitches have been detected in PSR J1833$-$0827, making it another frequent glitching pulsar. Some of the 16 glitches may experience exponential or linear recovery, but it is unlikely for us to make further analyses with the large gap in the data set.
All the glitch rates obtained from Nanshan are higher than that from Jodrell Bank Observatory. The small glitch size and high glitch rate could possibly attribute to the high observation cadence.
}
   \authorrunning{Z. R. Zhou et al. 2022}                     
   \titlerunning{Small glitches were detected at Nanshan}       

   \maketitle


\section{Introduction}\label{Sect.1}

Pulsars are considered to be remarkably stable rotators, which can be used to establish a pulsar time-scale \citep{Hobbs2020}, search for ultra-low frequency gravitational waves \citep{Hobbs2019, Shannon2017} and tests of general relativity \citep{Kramer2006}. These results all depended on the so-called pulsar timing technique \citep{Yu2013}.
Details of the pulsar timing technique are described in \citet{Edwards2006}. The critical quantity of the pulsar timing technique is the times-of-arrival (TOAs) of the observed pulses, which are always compared with predicted arrival times. The predicted arrival times are determined using a model of the pulsar’s position, rotation, etc. The differences between the predicted and the actual TOAs are known as the pulsar timing residuals \citep{Shaw2018}. For an ideal model, the timing residuals would be dominated by measurement errors and have a white spectrum \citep{Groth1975}. Any features observed in the timing residuals indicate the existence of unmodelled effects, which may include calibration errors, orbital companions, or spin-down irregularities \citep{Kerr2020}.

Two main types of timing irregularity have been recognized: timing noise and glitches. The physical processes behind both phenomena are not well understood. Timing noise is an unmodelled feature in the timing residuals relative to a simple slowdown model \citep{Cordes2010} and has been seen in all classes of pulsars \citep{Parthasarathy2019}. It can be described as a random walk of the residuals, sometimes exhibiting a clear quasi-periodic behavior \citep{Hobbs2010}. Some of these features have been shown to arise in instabilities in the pulsar magnetosphere, which result in state changes in the spin-down rate \citep{Lyne2010}. A glitch is a sudden increase in rotation frequency of a pulsar, and the observed fractional glitch size range from $\sim 10^{-12}$ to $\sim 10^{-5}$ with a bimodal distribution \citep{Yu2013}. Glitches are rare and unpredictable events and vary significantly for different pulsars \citep{Espinoza2011}. Most glitches have been observed in relatively young pulsars, but they have also been observed in magnetars and millisecond pulsars \citep{Kaspi2017}.
Pulsar sometimes enters a recovery process following a glitch, in which the rotation frequency decays towards the pre-glitch value. The glitch recovery can be interpreted as a signature of the presence of a superfluid in the interior of the star \citep{Baym1992}.
A better understanding of glitches will provide an insight into the interior structure of neutron stars (NS).
Glitches are thought to be triggered either by the sudden transfer of angular momentum from the faster-rotating crustal neutron superfluid to the rest of the NS \citep{Chamel2013} or by the NS crustquakes \citep{Bransgrove2020}. The exponential recoveries can be explained as the re-establishment of equilibrium between pinning and unpinning in a vortex-creep region interior to an NS \citep{Alpar1993}. More than fifty years after the discovery of the first glitch, the exact origin of these phenomena is not yet well understood \citep{Haskell2015}.

This paper presents the results of a new search for glitches performed using the Nanshan pulsar timing data base. Sixteen new glitches found in this work are used to study the glitching behavior of pulsars. We describe our observations in Section 2. Our results are shown in Section 3. We discuss the results and conclude the paper in Sections 4 and 5.

\section{Observation and Data Analysis}
\label{sect:Obs}

Timing observations of the nine pulsars at Xinjiang Astronomical Observatory (XAO) are carried out with the 25-m telescope at Nanshan \citep{Wang2001}. A room temperature receiver was used and then updated to a cryogenic system in July 2002. The receiver has a bandwidth of 320 MHz and is centered at 1540 MHz. Before 2010, an analog filter bank (AFB) with 128 × 2.5 MHz sub-channels was used to take data. After January 2010, the data were obtained by a digital filterbank system (DFB) with 1024 channels. These nine pulsars have been generally observed three times per month, and the presented timing data were collected between January 4th, 2000, and March 4th, 2014. Observation's integration times for each pulsar are from 4 to 16 minutes.

The {\sc psrchive} software package (\citealt{Hotan2004}) is used for offline data reduction. An initial set of parameters is obtained from the ATNF pulsar catalog \citep{Manchester2005}. After dedispersing and removing radio-frequency-interference (RFI), the data are summed in time, frequency, and polarization to produce a mean pulse profile. An analytic template is obtained by fitting single or multiple Gaussian components to the mean pulse profile. The template is then cross-correlated with each observation to obtain pulse TOAs. The {\sc tempo2} (\citealt{Hobbs2006}) package is then used to convert the topocentric TOAs to the arrival times at the solar system barycentre with the Jet Propulsion Laboratory (JPL) planetary ephemeris DE421 \citep{Folkner2009}. Each observed TOA was referred to terrestrial time (TT) as realized by International Atomic Time (TAI).

The pulse phase $\phi$ predicted by standard timing model can be expressed as:
\begin{equation}
\label{eq:phase}
 \phi(t) =\phi_{0}+\nu(t-t_{0})+\frac{1}{2}\dot{\nu }(t-t_{0})^{2}+\frac{1}{6}\ddot{\nu } (t-t_{0})^3,
\end{equation}
where $\phi_{0}$ is the pulse phase at the time $t_{o}$ ,$\nu$ ,$\dot{\nu}$ and $\ddot{\nu}$ are the rotational frequency and its derivatives respectively.

The additional pulse phase induced by a glitch is described by:
\begin{equation}
\label{eq:glitch}
\begin{split}
\phi _{\rm{g}}=\Delta \phi +\Delta \nu _{\rm{p}}(t-t_{\rm{g}})+\frac{1}{2}\Delta \dot{\nu } _{\rm{p}}(t-t_{\rm{g}})^{2}\\
+[1-e^{-(t-t_{\rm{g}})/\tau _{\rm{d}}}]\Delta\nu_{\rm{d}}\tau_{\rm{d}},
\end{split}
\end{equation}
where glitch is described by an offset in pulse phase $\Delta \phi$ and the permanent increments in the pulse frequency $\Delta \nu _{p}$ and its first frequency derivative $\Delta\dot{\nu}_{p}$ ,in addition to a temporary frequency increment $\Delta\nu_{d}$ which decays exponentially with a timescale $\tau_{d}$ .Glitch epoch $t_{g}$ was taken to be halfway between the last pre-glitch observation and the first post-glitch observation.
The changes in the pulse frequency and its first derivative at the glitch are described as
\begin{equation}
 \label{eq:phase}
 \Delta \nu _{\rm{g}}=\Delta \nu _{\rm{p}}+\Delta \nu _{\rm{d}}
\end{equation}
and
\begin{equation}
\label{eq:phase}
\Delta \dot{\nu } _{\rm{g}} = \Delta \dot{\nu } _{\rm{p}} - \frac{\Delta\nu_d}{\tau_{d}},
\end{equation} respectively. In addation, the degree of recovery can be described by:
\begin{equation}
\label{eq:phase}
Q = \Delta\nu_{d}/\Delta\nu_{g}
\end{equation}
The signature of a glitch in plots of timing residuals and frequency with epoch has been demonstrated in Figure 1 of \citet{Espinoza2011}. In this work, all glitches in this paper were detected by visual inspection of the phase residuals. Any feature looking similar to those in Figure 1 of \citet{Espinoza2011} of was considered as a glitch candidate and explored in detail.

\section{Results}
\label{sect:result}

\begin{figure}
  \centering
 \includegraphics[width=7cm,angle=0]{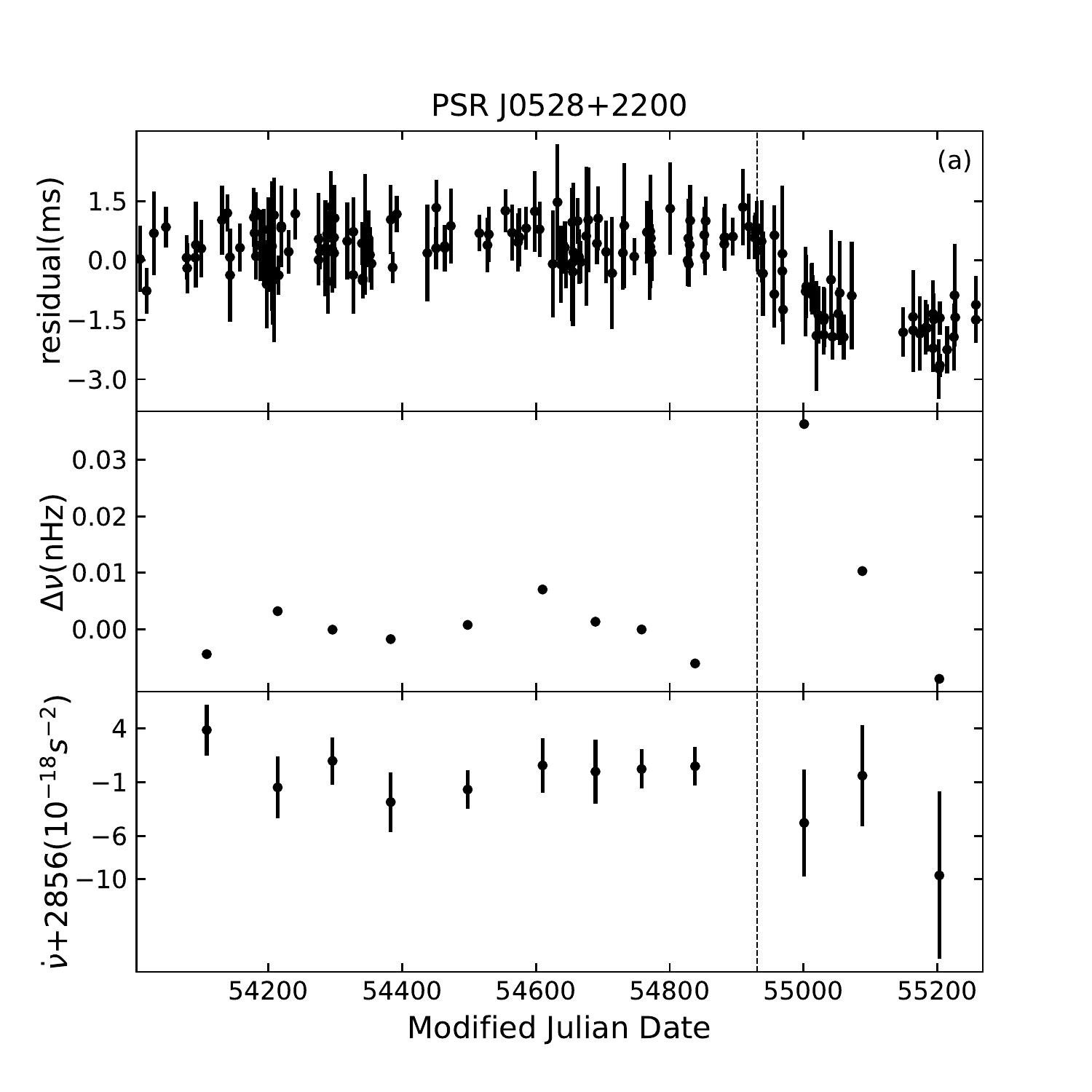}
 \includegraphics[width=7cm,angle=0]{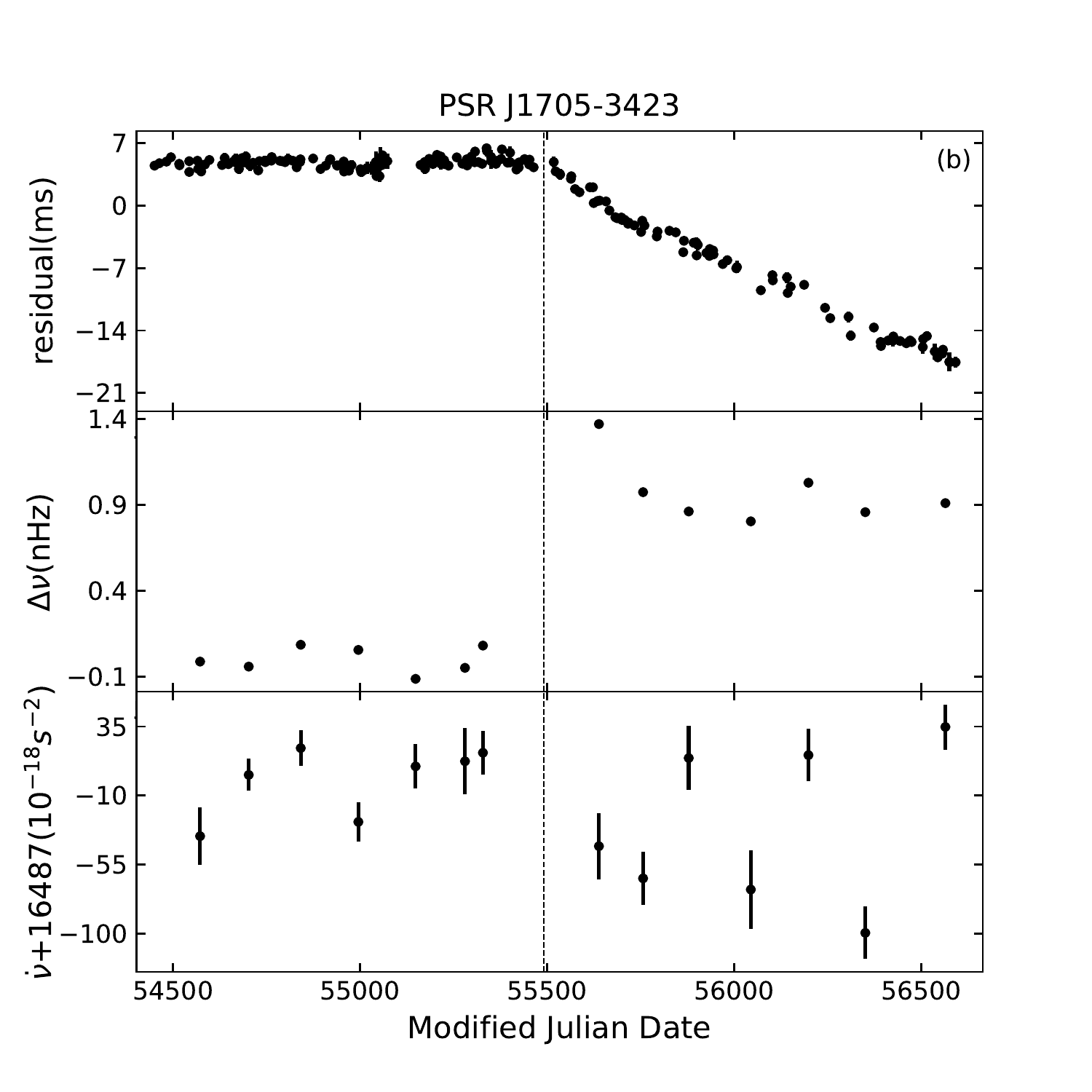}
 \includegraphics[width=7cm,angle=0]{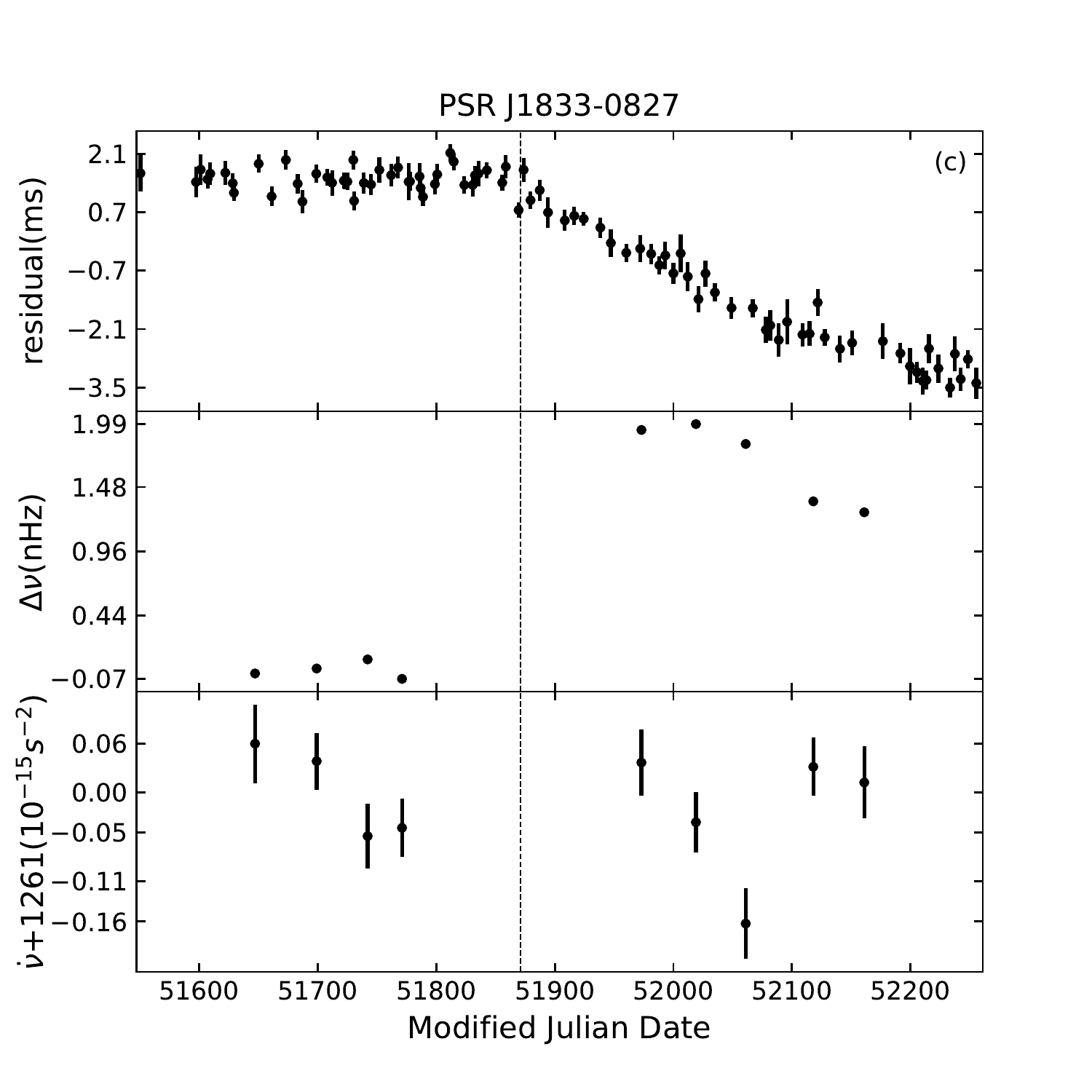}
 \includegraphics[width=7cm,angle=0]{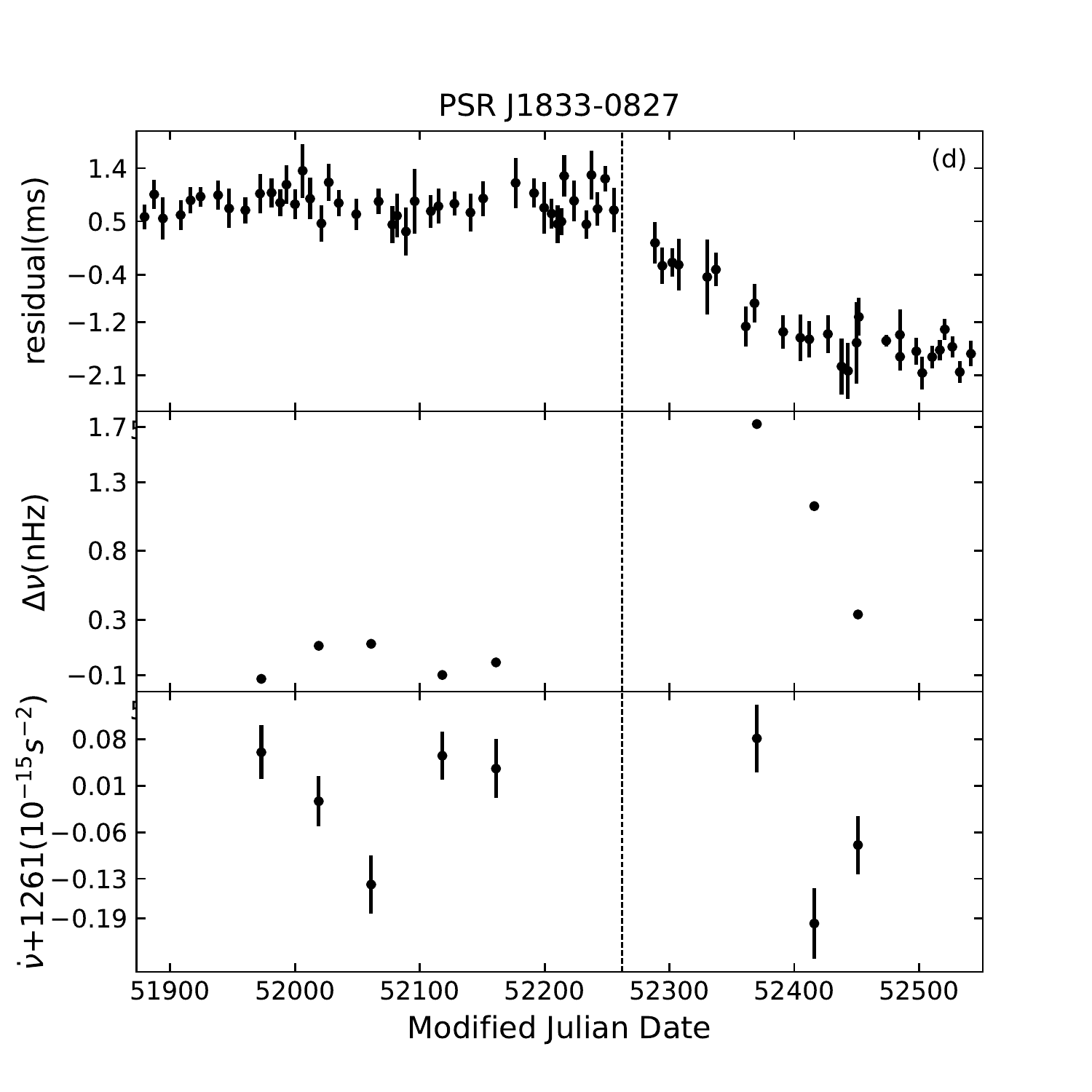}
 \includegraphics[width=7cm,angle=0]{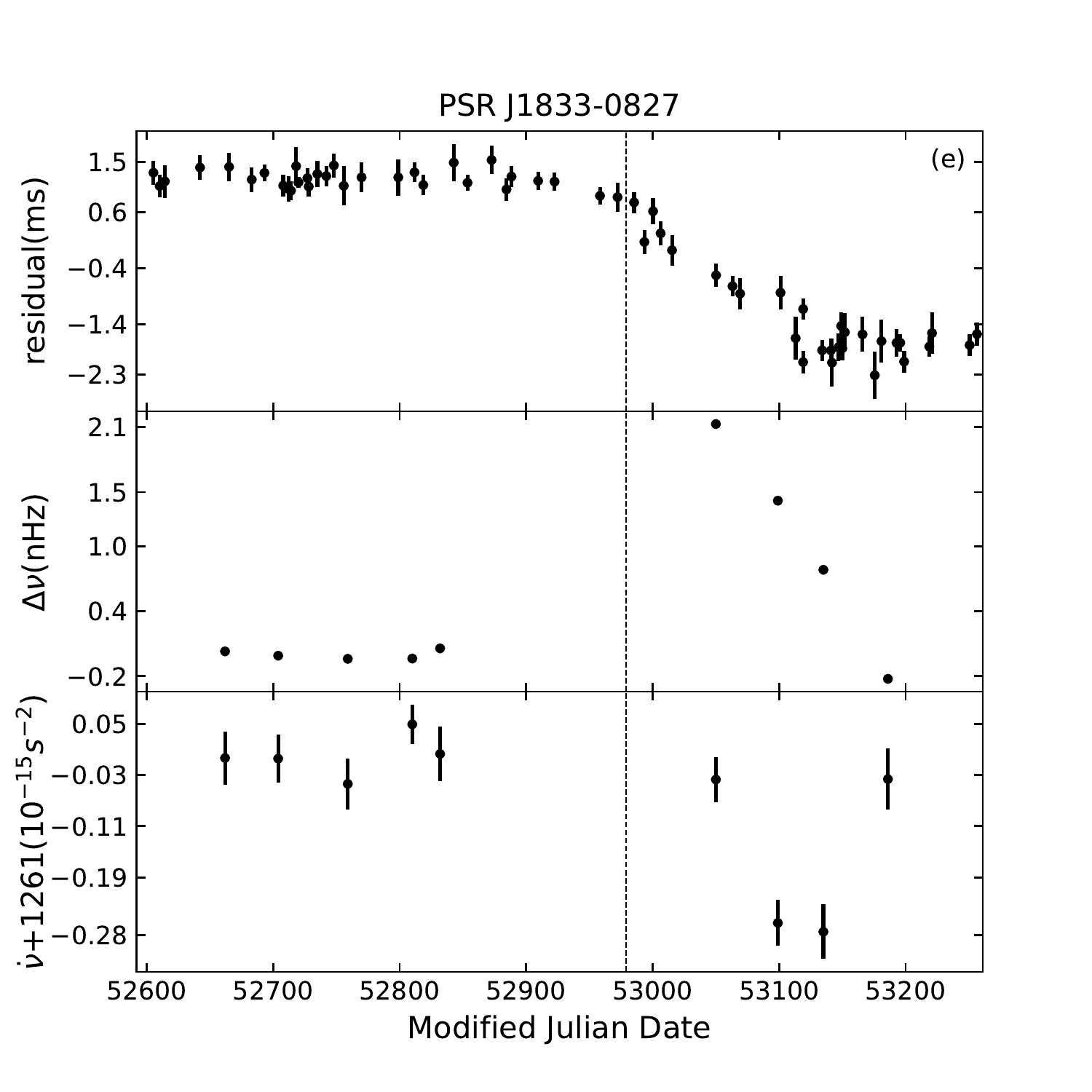}
 \includegraphics[width=7cm,angle=0]{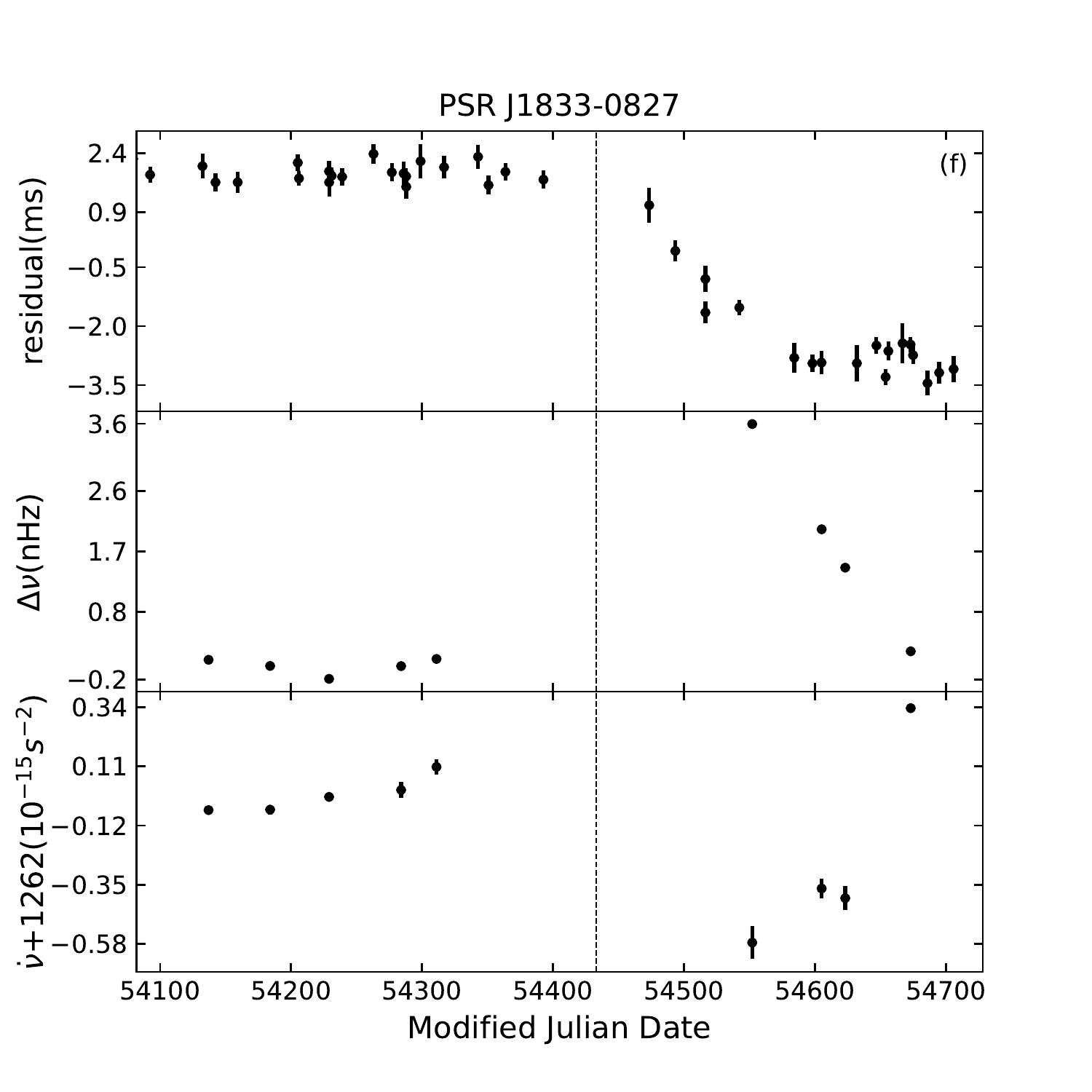}
 \caption{Sixteen glitches in nine pulsars. The top panel of each subplot shows timing residuals relative to the pre-glitch model; The middle panel is the variations of rotational frequency $\Delta\nu$ relative to the pre-glitch solutions; The bottom panel shows the variations of the pulse-frequency first-time derivative $\dot{\nu}$. The glitch epochs are indicated by vertical dashed lines within our data span.}\label{Fig1}
\end{figure}

 \begin{figure}
  \centering
  \includegraphics[width=7cm,angle=0]{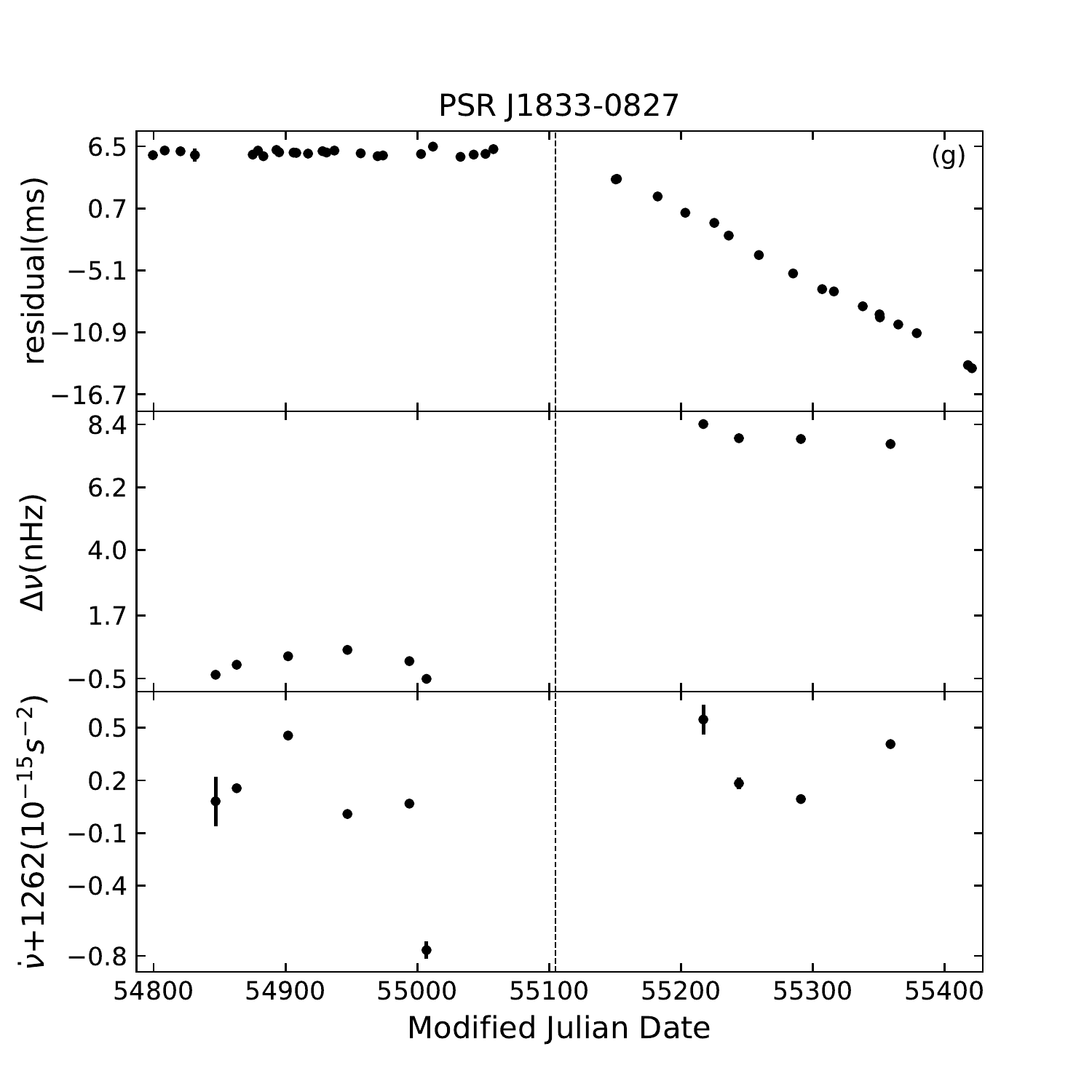}
  \includegraphics[width=7cm,angle=0]{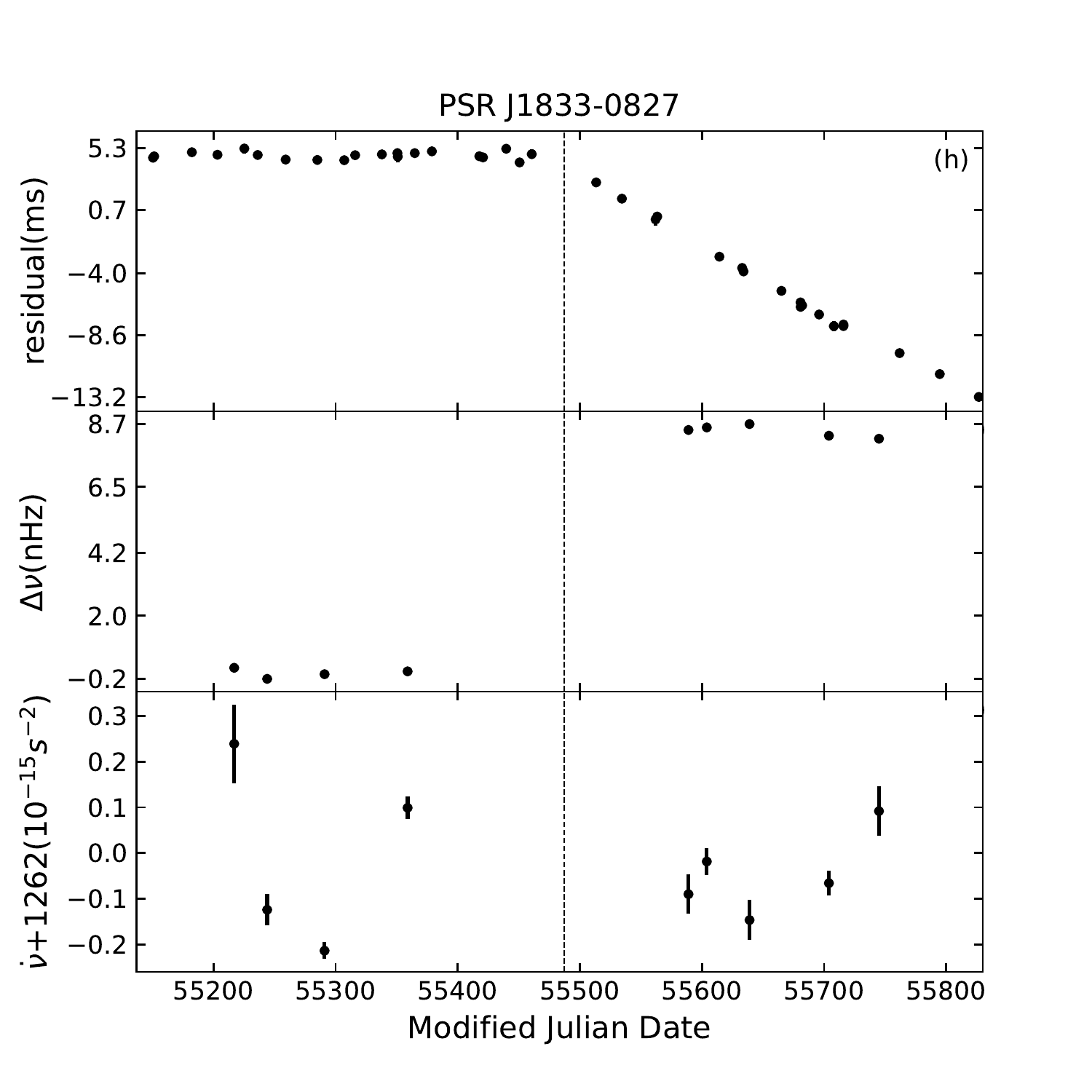}
  \includegraphics[width=7cm,angle=0]{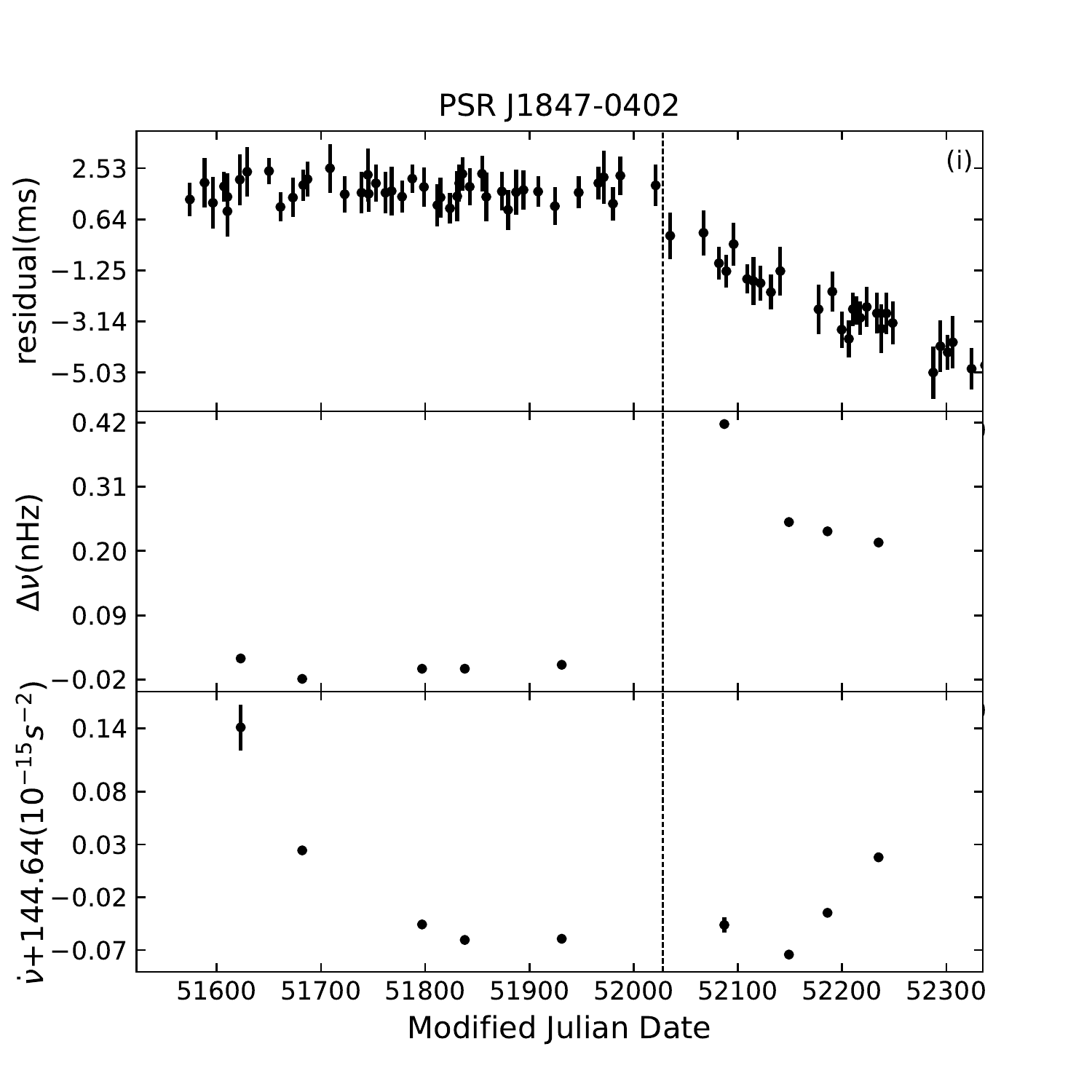}
  \includegraphics[width=7cm,angle=0]{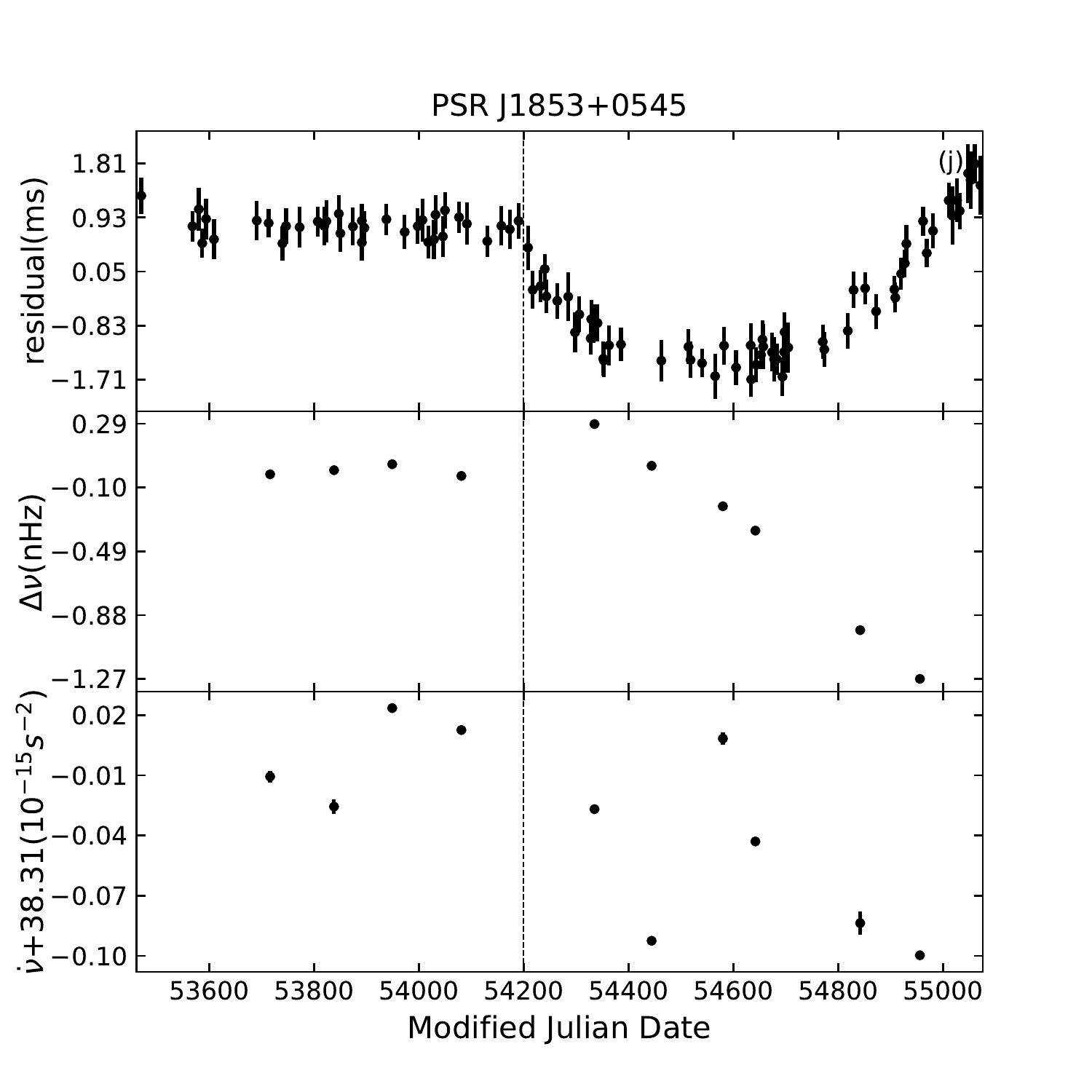}
  \includegraphics[width=7cm,angle=0]{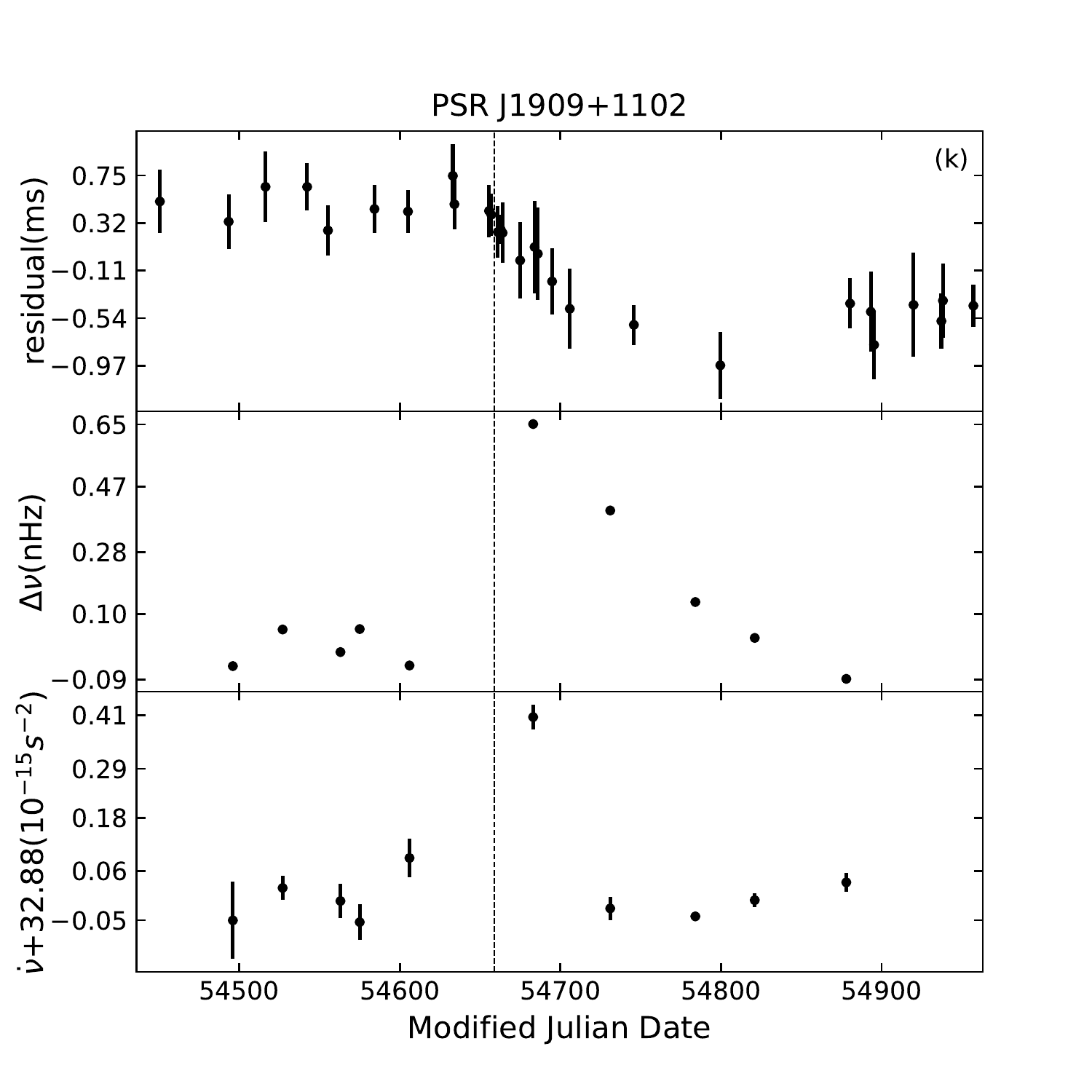}
  \includegraphics[width=7cm,angle=0]{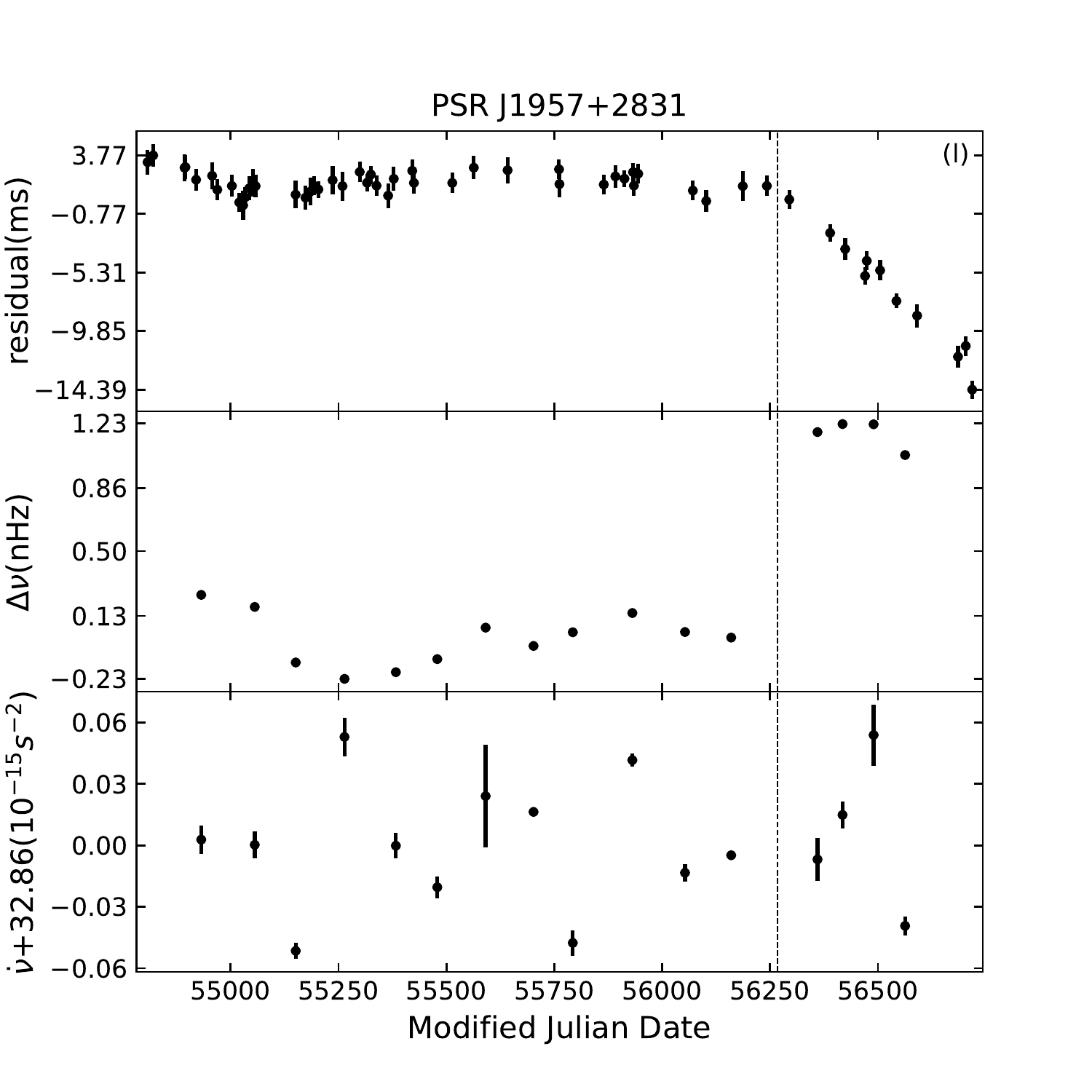}
  \addtocounter{figure}{-1}
  \caption{- continued}
 \end{figure}
 
 \begin{figure}
  \centering
  \includegraphics[width=7cm,angle=0]{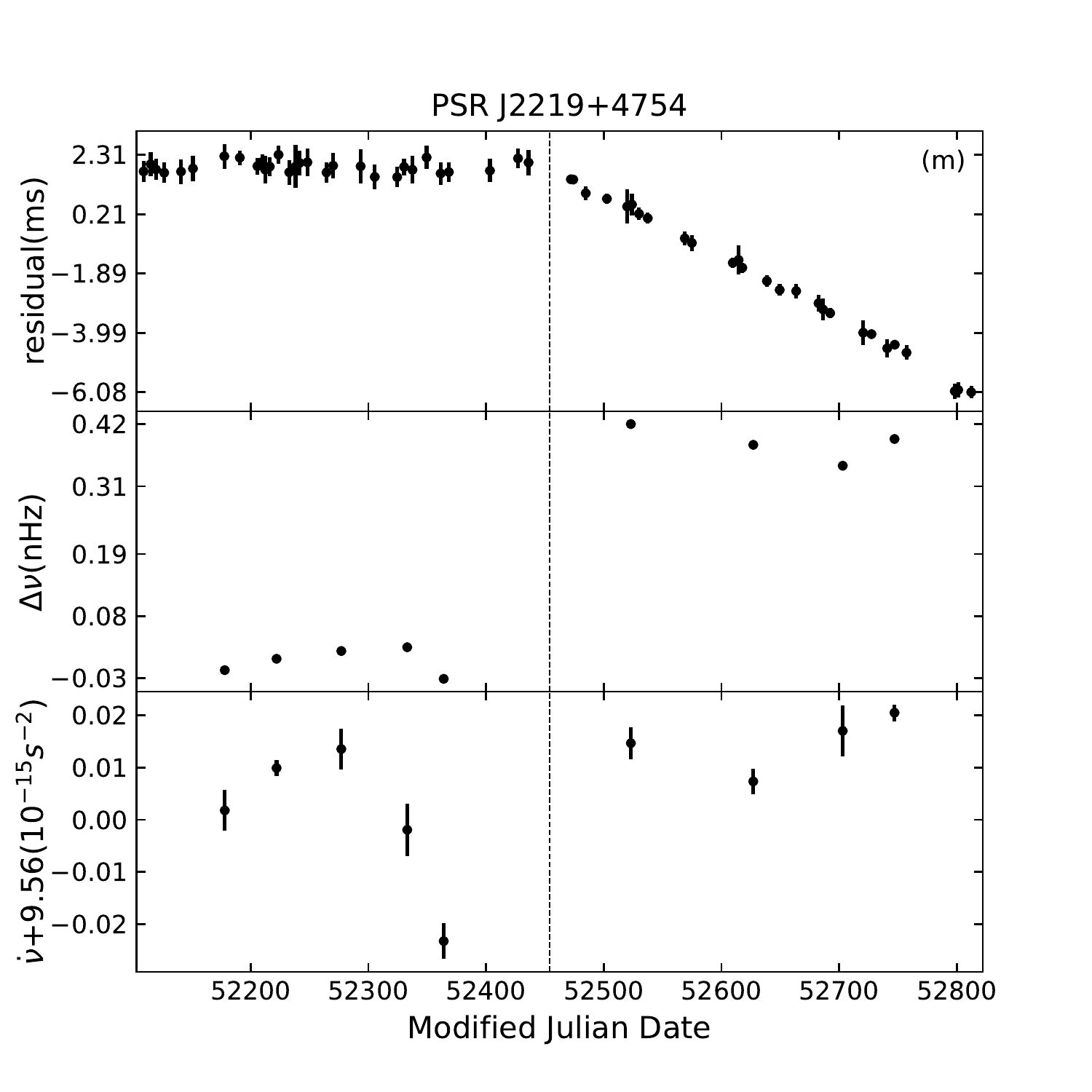}
  \includegraphics[width=7cm,angle=0]{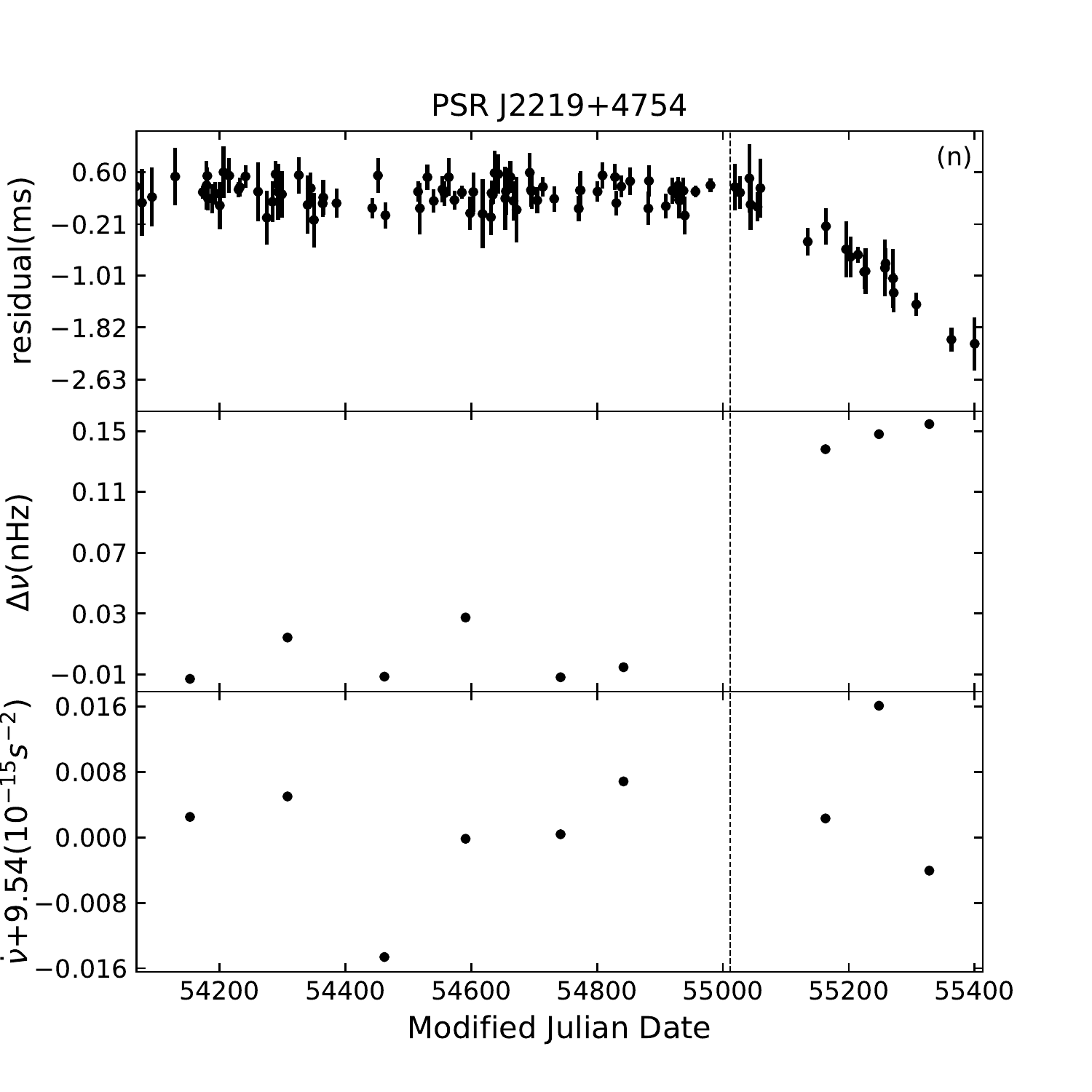}  
  \includegraphics[width=7cm,angle=0]{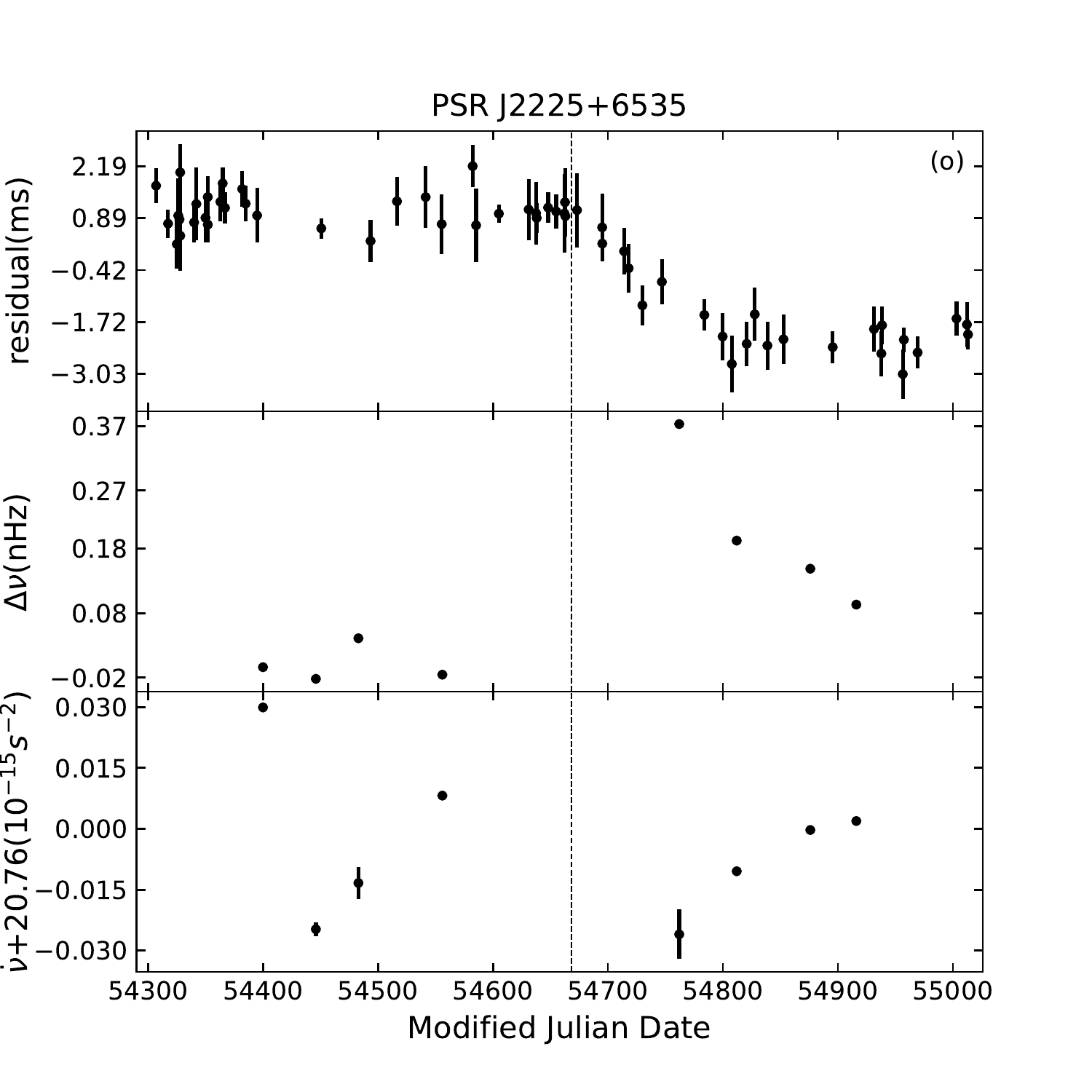}
  \includegraphics[width=7cm,angle=0]{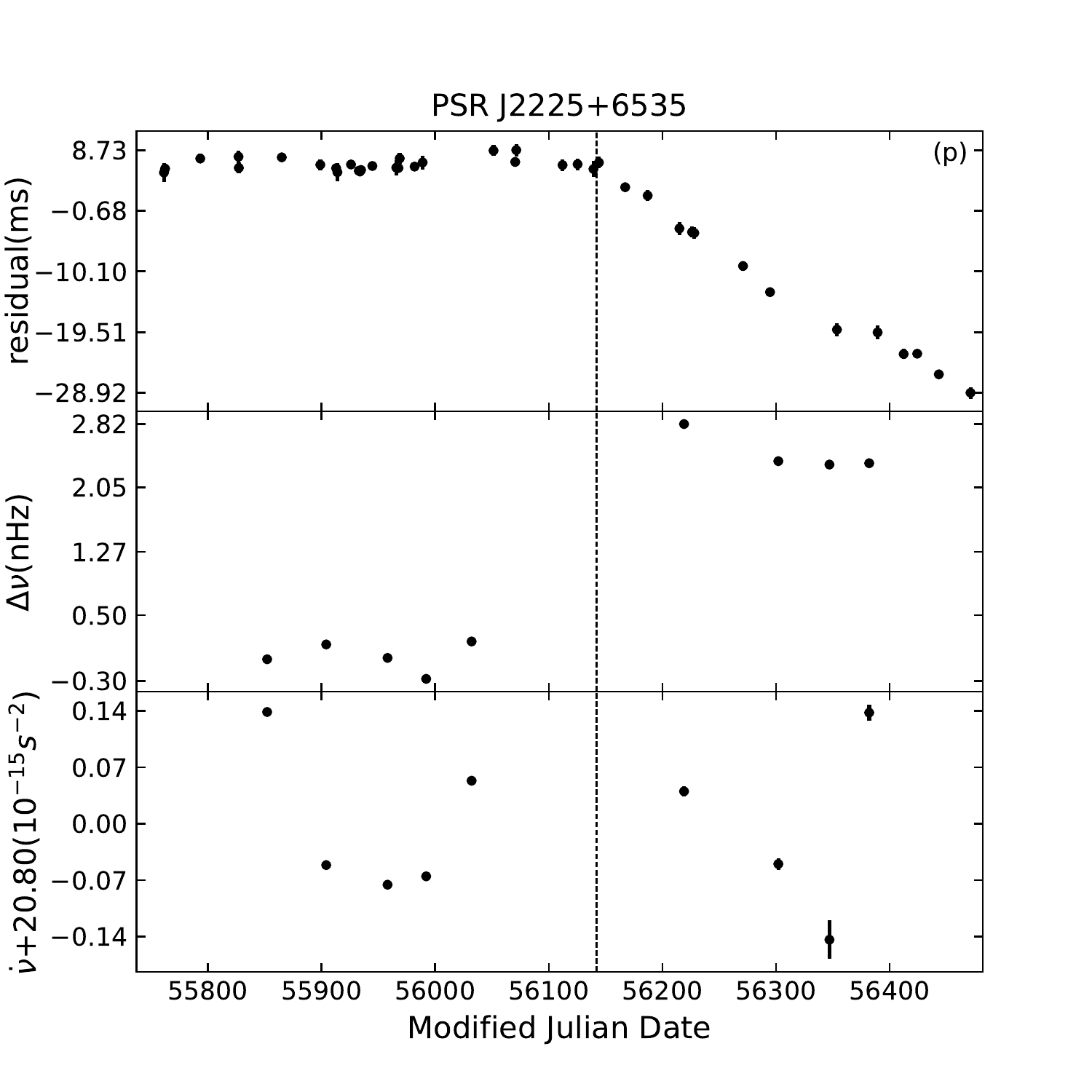}
  \addtocounter{figure}{-1}
  \caption{- continued}
 \end{figure}

\begin{table}
\bc
\begin{minipage}[]{100mm}
\caption[]{Glitch parameters from all of the new glitches in our sample. \label{tab1}}\end{minipage}
\setlength{\tabcolsep}{1pt}
\small
 \begin{tabular}{ccccccccc}
  \hline\noalign{\smallskip}
 Pulsar & Gl. No.  &   Epoch & $\Delta\nu/\nu$& $ \Delta\dot{\nu}/\dot{\nu} $ & Rms res. &  Data span &  No. of  & Age \\
 Name   & $\qquad$ &   (MJD) & ($10^{-9}$)   & ($10^{-3}$)  & ($\mu$s) &  (MJD) &  TOAs & (Myr) \\
  \hline\noalign{\smallskip}
 J0528+2200 & 1  & 54931.0(3) & 0.21(4) & 1.0(2)  & 488.634 & 54007$-$55263 & 160 &1.48\\
 J1705$-$3423 & 1  & 55490(13)  & 0.28(1) & 0.11(7) & 636.888  & 54450$-$56718&197& 3.76\\
 J1833$-$0827 & 1  & $51871(1)$  & $0.22(2)$& $0.02(1)$ & 134.933 & 51550$-$52262 & 65 & 0.147\\
 $\qquad$   & 2  & $52262(3)$  & $0.21(3)$ & $0.11(2)$ & 175.224 & 51872$-$52555 & 60 &\\ 
 $\qquad$   & 3  & $52979(3)$  & $0.29(2)$ & $0.17(1)$& 117.119 & 52555$-$53273 & 49 &\\
 $\qquad$   & 4  & $54433(20)$ & $0.61(5)$ & $0.29(3)$& 134.758 & 54058$-$54782 & 28 &\\
 $\qquad$   & 5 & 55105(23) & 0.64(1) & $-0.010(8)$ & 42.826 & 54782$-$55461 & 18 &\\
  $\qquad$   & 6 & 55487(13) & 0.64(2) &  0.06(1) & 97.024 & 55105$-$55859 & 27 &\\
 J1847$-$0402 & 1  & $52028(3)$  & $0.09(2)$& $-0.06(1)$  & 62.157 & 51550$-$52300 & 24 &0.183\\
 J1853$+$0545 & 1  & $54199(4)$  & $0.109(5)$ & $0.70(3)$ & 78.022 & 53500$-$55110 & 53 &3.27\\
 J1909$+$1102 & 1  & 54659(1)  & 0.12(2)   & 1.23(2) & 68.666 & 54400$-$54958 & 21 &1.70\\
 J1957$+$2831 & 1  & 56268(12) & 0.37(9)    & 0.06(4)  & 789.092 & 54800$-$56718 & 54& 1.57\\
  J2219$+$4754  & 1 & 52454(9)  & 0.18(1)    &  0.4(1)      & 64.625 & 52102$-$52874 & 37 &3.09\\
 $\qquad$   & 2  & 55012(3)  & 0.068(3)   &$-0.08(3)$    & 38.938 & 54004$-$55496 & 44 &\\
 J2225$+$6535 & 1  & $54668(3)$  & $0.30(1)$& $1.57(7)$ & 63.626 & 54300$-$55028 & 22 &1.12\\
 $\qquad$   & 2  & $56142(1)$  & $1.65(2)$ & $1.54(7)$ & 54.965 & 55761$-$56487 & 13 &\\
  \noalign{\smallskip}\hline
\end{tabular}
\ec
\end{table}

We have found $16$ new glitches in $9$ pulsars. Compared with \citet{Basu2022} and \citet{Lower2021}, most glitches detected in this work have small fractional size ($\Delta\nu_{g}/\nu \leq 10^{-9}$). Table 1 contains the parameters of each detected glitch. The first two columns list the pulsar name and the reference number for each detected glitch. The third column indicates the glitch epoch and its uncertainty in parentheses. The fourth and the fifth columns show the relative changes in the rotation frequency and its first-time derivative. The root means square of
residuals, data span, number of TOAs, and characteristic ages are separately given
in the rest of the columns. Except for the glitch epoch, all uncertainties are quoted at the $1\sigma$ level returned by {\sc tempo2}. The glitch epoch was kept halfway between the last pre-glitch observation and the first post-glitch observation, with an uncertainty of a quarter of the observation gap. In table 1, all these pulsars have been observed with glitches before \citep{Basu2022, Lower2021, Yu2013, Espinoza2011, Yuan2010}. There are 10 glitches reported in \citet {Lower2021} and \citet{Basu2022} were also detected by us. It is not necessary to repeat their results here.

\subsection{PSR J0528$+$2200 (B0525$+$21)}
This pulsating radio source was found near the Crab Nebula \citep{Staelin1968}. It is a very long-period pulsar with $P=3.746$ s. In total, four small glitch events had been reported for this pulsar \citep{Downs1982, Shemar1996, Janssen2006, Yuan2010}. We identified a new small glitch at MJD$\sim$54931 with a fractional size of $\Delta\nu/\nu = 0.21 \times 10^{-9}$, which is comparable with the glitch reported by \citet{Janssen2006} at MJD 53379. It is possible that a linear decrease of $\nu$ relative to the pre-glitch value lasts for at least 200 days and overshoots the pre-glitch extrapolation. This is similar to the behavior seen in the Crab pulsar \citep{Lyne1993}. Fig.1(a) shows the spin-down rate $|\dot{\nu}|$ of post-glitch is larger than the pre-glitch.

\subsection{PSR J1705$-$3423 (B1705$-$3423)}
PSR J1705$-$3423 was discovered in the Parkes Southern Pulsar
Survey ~\citep{Manchester1996}. It has a period of
255 ms and period derivative of $1.076\times 10^{-15}$, which implies
a characteristic age of 3.76 Myr. Two small glitches had
been detected by ~\citet{Espinoza2011} and ~\citet{Yuan2010} at MJD $\sim$ 51956 and MJD $\sim$ 54408, respectively. Another new small glitch was detected at MJD$\sim$55490 with a magnitude of $\Delta\nu \sim1.1\times 10^{-9}$ Hz. Noted that it is the smallest of the three glitches reported for this pulsar. Fig.1(b) presents the evolution of $\Delta\nu$ and $\dot{\nu}$ of this pulsar for 6.2 years. There is a possible exponential recovery with a time constant of about 400 d. There is some evidence that the fluctuation of $\dot{\nu}$ after the glitch is greater than that before.

\subsection{PSR J1833$-$0827 (B1830$-$08)}
This pulsar was discovered in a high-radio-frequency survey for distant pulsars carried out at Jodrell Bank~\citep{Clifton1986}. PSR J1833$-$0827 has a period of 85 ms and period derivative of $9.18\times 10^{-15}$ giving it a young characteristic age of 147 kyr. This pulsar suffered a giant glitch of fractional size $\Delta\nu/\nu=1.9\times10^{-6}$ and $ \Delta\dot{\nu}/\dot{\nu}=1.7\times10^{-3}$ in June 1990~\citep{Shemar1996}.~\citet{Espinoza2011} reported another one with an amplitude of $\Delta\nu/\nu=0.9\times10^{-9}$ at MJD$\sim$47541. Six new small glitches have been detected in this pulsar after MJD $\sim$ 48051, which makes it another frequent glitching pulsar. The relative changes of the rotational frequency $\Delta\nu/\nu$ shown in Fig.s 1(c)$\sim$1(h) ranges from $0.21\times10^{-9}$ to $0.64\times10^{-9}$. 
Fig.1(c) and Fig.1(d) show the spin-down rate $|\dot{\nu}|$ of post-glitches are larger than the pre-glitches. There is likely a linear decrease of $\nu$ relative to the pre-glitch and overshoots the pre-glitch extrapolation in Fig.1(e) and Fig.1(f), while $\Delta\nu$ is almost unchanged in Fig.1(g) and Fig.1(h) after glitches. A gap lasting for $\sim$ 100 days exists in glitches shown in Fig.1(f) and Fig.1(g). Besides, both Fig.1(e) and Fig.1(f) show the $\dot{\nu}$ fluctuations of post-glitches are greater than pre-glitches.  

\subsection{PSR J1847$-$0402 (B1844$-$04)}
PSR J1847$-$0402 was discovered in a systematic search at low galactic latitudes near the center frequency of 408 MHz using the Mark I radio telescope at Jodrell Bank \citep{Davies1970}. This pulsar has a period of 598 ms and a modest period derivative ($\dot{P}\sim 5.17\times10^{14}$), implying a young pulsar with a characteristic age of about 183 kyr. Two small glitches were detected for PSR B1844$-$04 around MJD $\sim$ 55502 and 58244 with $\Delta\nu/\nu\sim 10^{-10}$ ~\citep{Basu2022,Lower2021}. Fig.1(i) shows another new small glitch at MJD$\sim$52028. There is seemingly an exponential recovery of a small part of the change in $\nu$ with a time constant of about 60 d. 

\subsection{PSR J1853$+$0545 (B1853$+$0545)}
PSR J1853+0545 was discovered in the Parkes Multi-beam Pulsar Survey-\textrm{III}~\citep{Kramer2003}. It has a period of 126 ms and a small period derivative($\dot{P}\sim6.12\times10^{-16}$), implying a relatively large characteristic age of 3.27 Myr. \citet{Yuan2010} detected the first known glitch with a small magnitude of $\Delta\nu/\nu=1.46\times10^{-9}$, following an exponential recovery with fractional decay $Q=0.22$. 
Fig.1(j) presents another new small glitch measured at MJD $\sim$ 54199. Similar to PSR J0528$+$2200, there is possibly a linear decrease of $\nu$ relative to the pre-glitch value that lasts for at least 600 days and overshoots the pre-glitch extrapolation. Once again, this is similar to the Crab pulsar post glitch behavior \citep{Lyne1993}. The bottom panel of Fig.1(j) shows a noticeable fluctuation in $\dot{\nu}$ after the glitch.

\subsection{PSR J1909$+$1102 (B1907$+$10)}
PSR J1847$-$0402 was discovered in a systematic search at low galactic latitudes, using the Mark I radio telescope and a radio frequency of 408 MHz \citep{Davies1973, Davies1972}.
PSR B1907+10 has a period of 284 ms and is relatively old. \citet{Yuan2010} found a small glitch at MJD $\sim$ 52700 with $\Delta\nu/\nu\sim0.27\times10^{-9}$. Following that, there exists another glitch detected by us at MJD $\sim$ 54659, with the magnitude of $0.12\times10^{-9}$. The post-glitch behaviors of $\Delta\nu$ and $\dot{\nu}$ are shown in Fig.1(k) and are similar to that of the glitch in Fig.1(j). The spin-down rate $|\dot{\nu}|$ of after the glitch is larger than that before. The fluctuation of $\dot{\nu}$ before the glitch is smaller than that after. 

\subsection{PSR J1957$+$2831}
PSR J1957+2831 was found during the search of the supernova remnants G65.1$+$0.6 using the 76-m Lovell radio telescope at Jodrell Bank \citep{Lorimer1998}. It has a period of 308 ms and characteristic age $\tau_{c}\sim1.57 $ Myr. A new small glitch with $\Delta\nu/\nu \sim0.37\times10^{-9}$ was observed at MJD $\sim$ 56278. This glitch is the fourth jump event in frequency for PSR J1957$+$2831 after three glitches were reported by \citet{Espinoza2011}. Fig.1(l) shows that no significant changes in spin-down rate ($|\dot{\nu}|$) can be associated with this glitch. This event leads to an increase in $\nu$ with $\Delta\nu \sim$ 1.5 nHz. The fluctuation of $\dot{\nu}$ is similar before and after the glitch.

\subsection{PSR J2219$+$4754 (B2217$+$47)}
This pulsar was discovered during a search for such objects carried out with the 300-foot transit telescope at the US National Radio Astronomy Observatory in October and November 1968 \citep{Taylor1969}. It is a slow pulsar with a period of 538 ms, and \citet{Michilli2018} had presented a comprehensive study of profile variations in PSR B2217$+$47. PSR J2219$+$4754 is relatively old ($\tau_{c}\sim3.09 $ Myr). \citet{Basu2022} observed a glitch at MJD$\sim$55857 with magnitude $\Delta\nu/\nu \sim1.16\times 10^{-9}$. We measured two new glitches in this work. Fig.1(m) shows the same trend between frequency increment and spin-down rate of post-glitch. 
There is probably a continuous increase in the frequency relative to the extrapolated pre-glitch that lasts for at least 300 days, as shown in Fig.1(n). This is unusual for post-glitch behavior. Such behavior has also been seen for PSR J0147$+$5922 at MJD$\sim$53682 \citep{Yuan2010}.

\subsection{PSR J2225$+$6535 (B2224$+$65)}
PSR J2225+6535 was found in a low latitude pulsar survey using the Mark 1A radio telescope at Jodrell Bank in 1972 \citep{Davies1973}. This pulsar is a slow-spin ($P\sim 683$ms), high-velocity pulsar associated with the Guitar Nebula \citep{Cordes1993}. It seems to be a young pulsar with a characteristic age of about $1.1$ Myr. We measured two small glitches after five glitches detected by~\citet{Backus1982},~\citet{Janssen2006} and~\citet{Yuan2010}, respectively. Among the five previous glitches, only the first one is a large glitch with $\Delta\nu/\nu \sim 1.7 \times 10^{-6}$ at MJD$\sim$43072~\citep{Backus1982}. As seen in Fig.1(o), there is the possibility of a short-term ($\sim$50 d) exponential recovery in $\dot{\nu}$. Although the post-glitch data span in Fig.1(p) is short, it is likely that there is an exponential recovery of a small part of the change in frequency after the jump, but it can not recover to the trend before the glitch. This behavior is very similar to that observed in the Vela pulsar~\citep{Lyne1996}. Compared to the pre-glitch, no or little change in post-glitch $\dot{\nu}$ was observed.

\section{Discussion}
\label{sect:discussion}

\begin{table}
\begin{center}
\caption[]{Observation times span and glitch rate of 9 pulsars observed at Jodrell Bank Observatory (JBO) or Nanshan have glitched. The beginning and end data of JBO refer to \citet{Espinoza2011} and \citet{Basu2022} respectively.}\label{Tab1}
 \begin{tabular}{ccccccc}
  \hline\noalign{\smallskip}
   PSR J &\multicolumn{2}{c}{Range (MJD)}&\multicolumn{2}{c}{No. of glitches}&\multicolumn{2}{c}{Glitch rate ($R_{g}$, yr$^{-1}$)}\\
    & JBO & Nanshan & JBO & Nanshan & JBO & Nanshan \\
  \hline\noalign{\smallskip}
  J0528$+$2200&45010$-$58482&51547$-$56719&4&4&0.11(5)&0.28(14)\\
  J1705$+$3423&49086$-$58482&52485$-$56719&3&2&0.12(7)&0.17(12)\\
  J1833$-$8273&46449$-$58482&51549$-$56719&3&7&0.09(5)&0.49(19)\\
  J1847$-$0402&44816$-$58482&51550$-$56719&3&3&0.08(5)&0.21(12)\\
  J1853$+$0545&51634$-$58482&52497$-$56719&1&2&0.05(5)&0.17(12)\\
  J1909$+$1102&44816$-$58482&52470$-$56719&1&2&0.03(3)&0.17(12)\\
  J1957$+$3831&50239$-$58482&52503$-$56719&3&3&0.13(8)&0.26(15)\\
  J2219$+$4754&45953$-$58482&51549$-$56719&1&3&0.03(3)&0.21(12)\\
  J2225$+$6535&44817$-$58482&52470$-$56719&4&5&0.11(5)&0.43(19)\\
 \noalign{\smallskip}\hline
\end{tabular}
\end{center}
\end{table}

\begin{figure}
  \centering
  \includegraphics[width=7.1cm, angle=0]{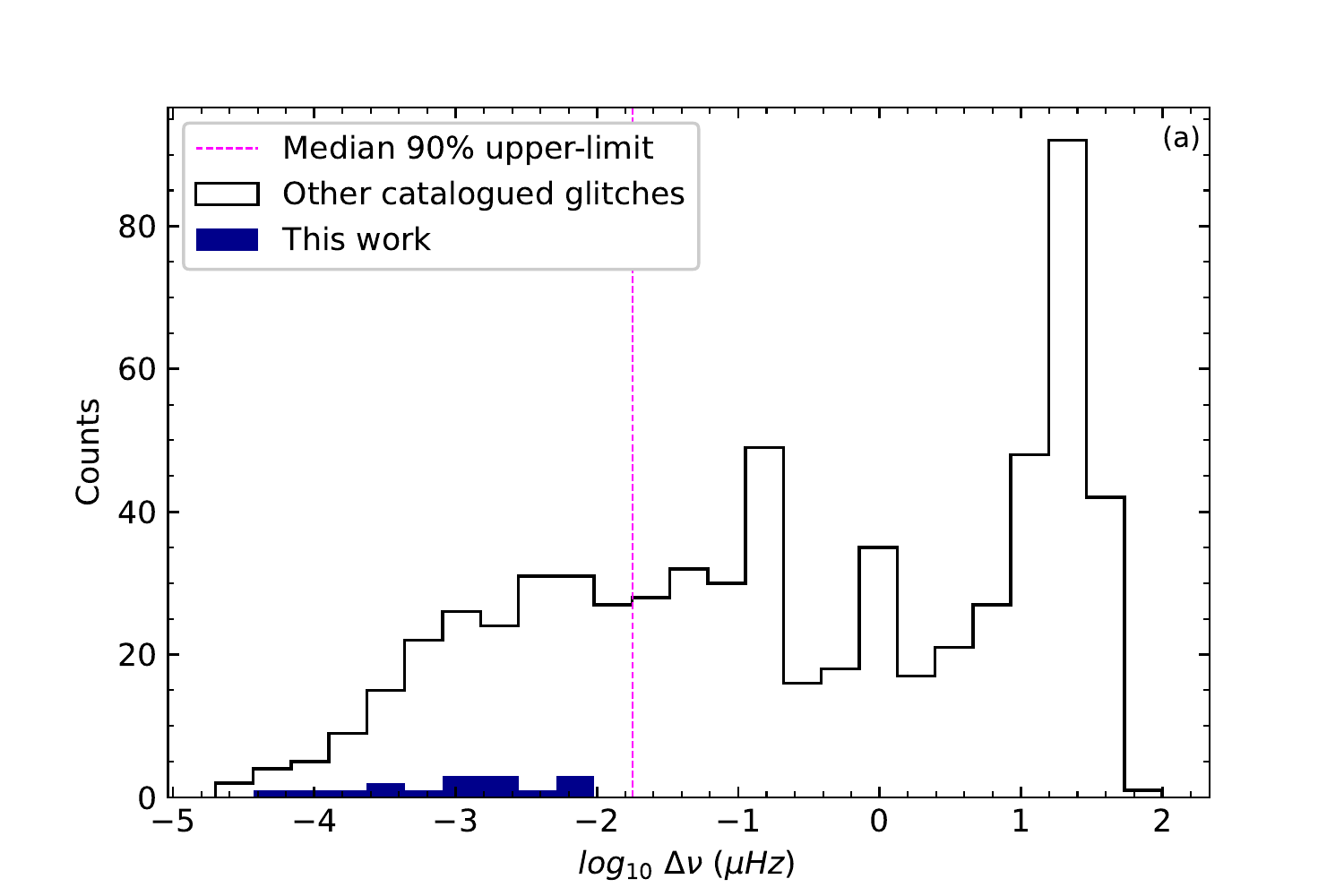}
  \includegraphics[width=7.1cm, angle=0]{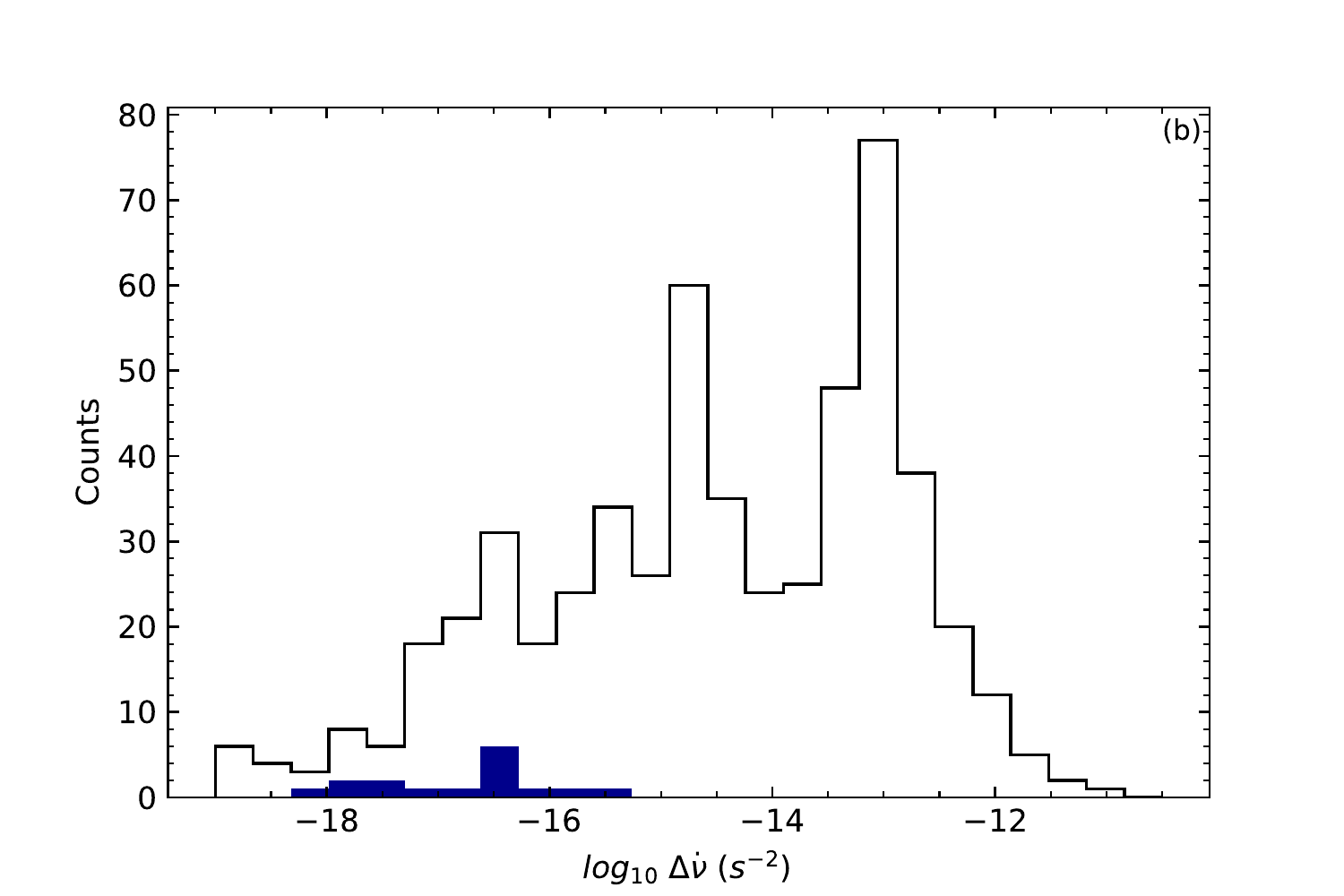}
  \includegraphics[width=7.1cm, angle=0]{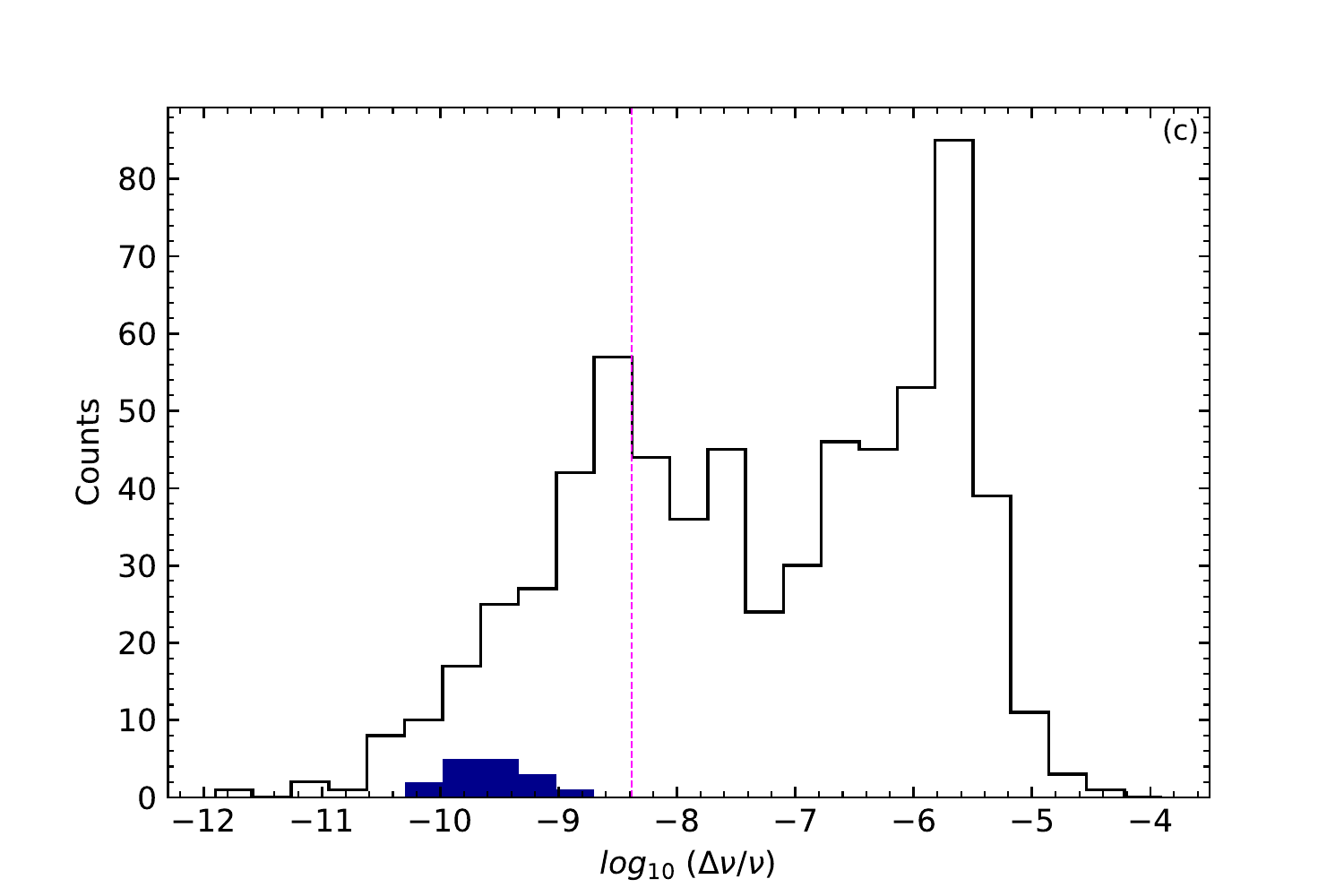}
  \includegraphics[width=7.1cm, angle=0]{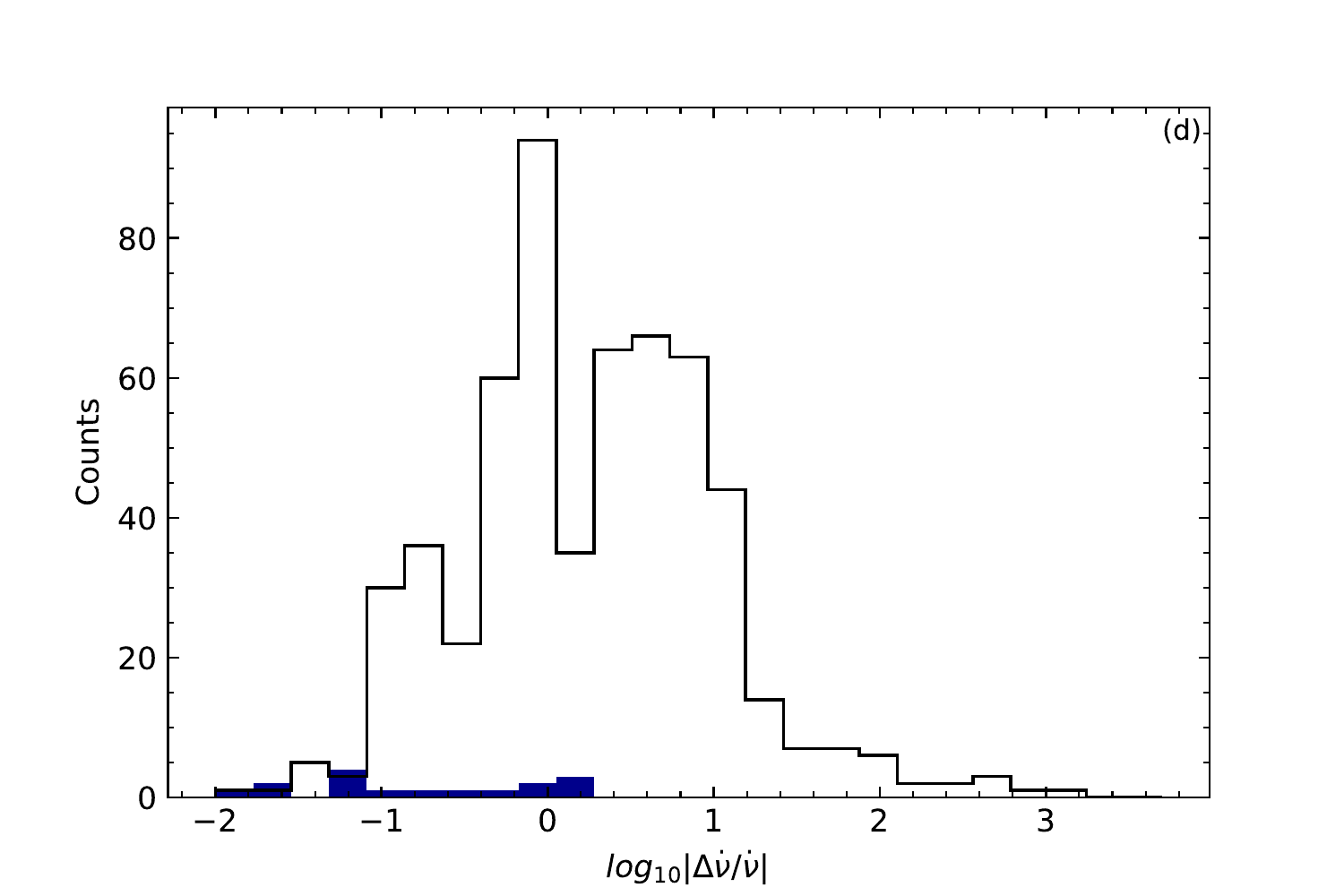} 
\caption{Histograms of $\Delta\nu$, $\Delta\Dot{\nu}$, $\Delta\nu/\nu$, and $\Delta\dot{\nu}/\dot{\nu}$ for the 16 glitches in our sample (dark blue) and the JBO
glitch catalogue after removing those listed in Table 1 (grey). The dashed magenta lines in panels (a) and (c) indicate the median upper limit on the glitch size after averaging across our sample.}\label{abcd} \end{figure}

 \begin{figure}
 \centering
 \includegraphics[width=1\textwidth, angle=0]{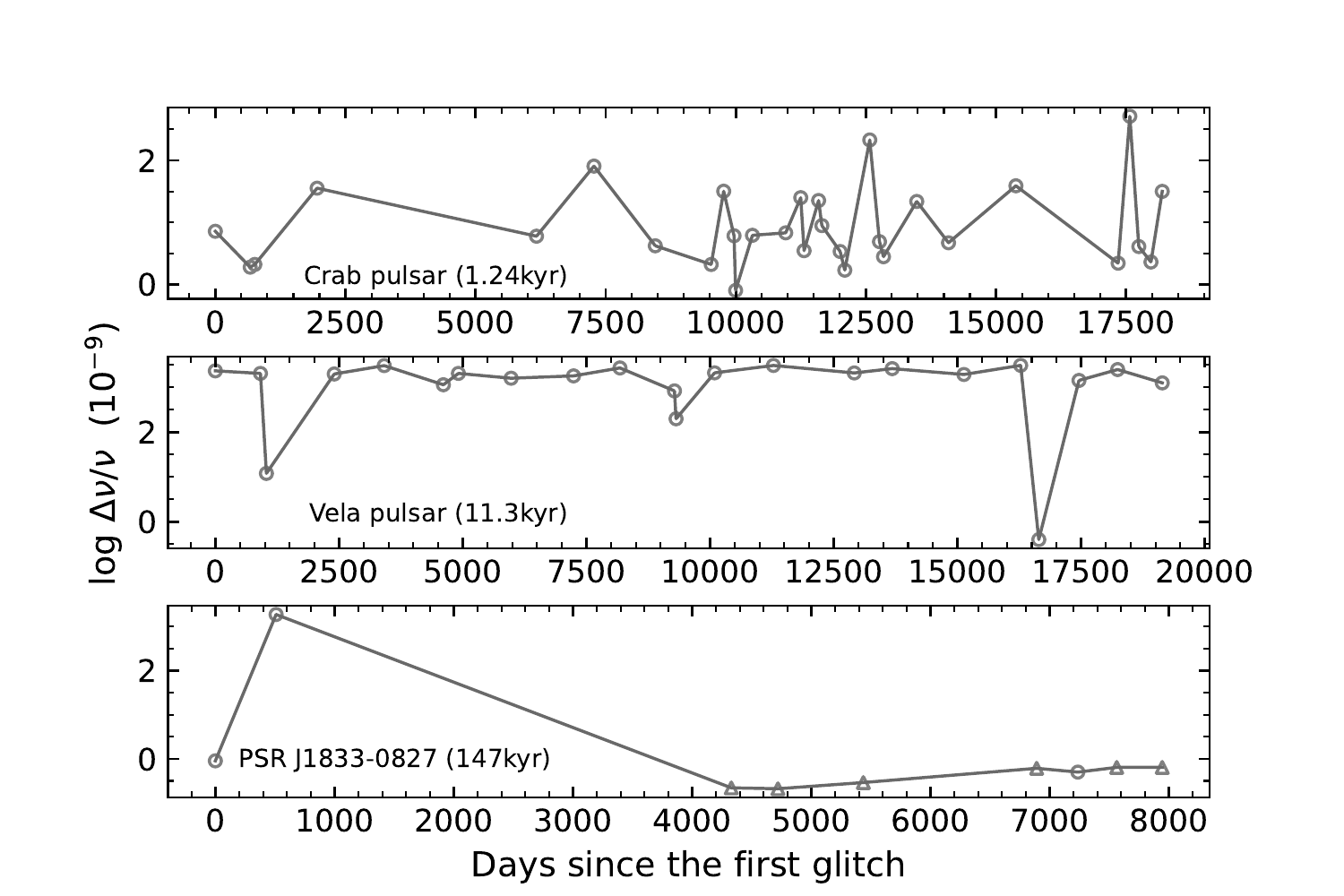} 
 \caption{The time series of glitch events for PSR J1833$-$0827, Crab and Vela pulsars. For each pulsar, its characteristic age is shown in parentheses. Data from the JBO pulsar catalogue glitch table (http://www.jb.man.ac.uk/pulsar/glitches.html) are indicated by circles, while those from this work are indicated by triangles.}
 \label{3psr}
 \end{figure}

 \begin{figure}
 \centering
 \includegraphics[width=0.8\textwidth, angle=0]{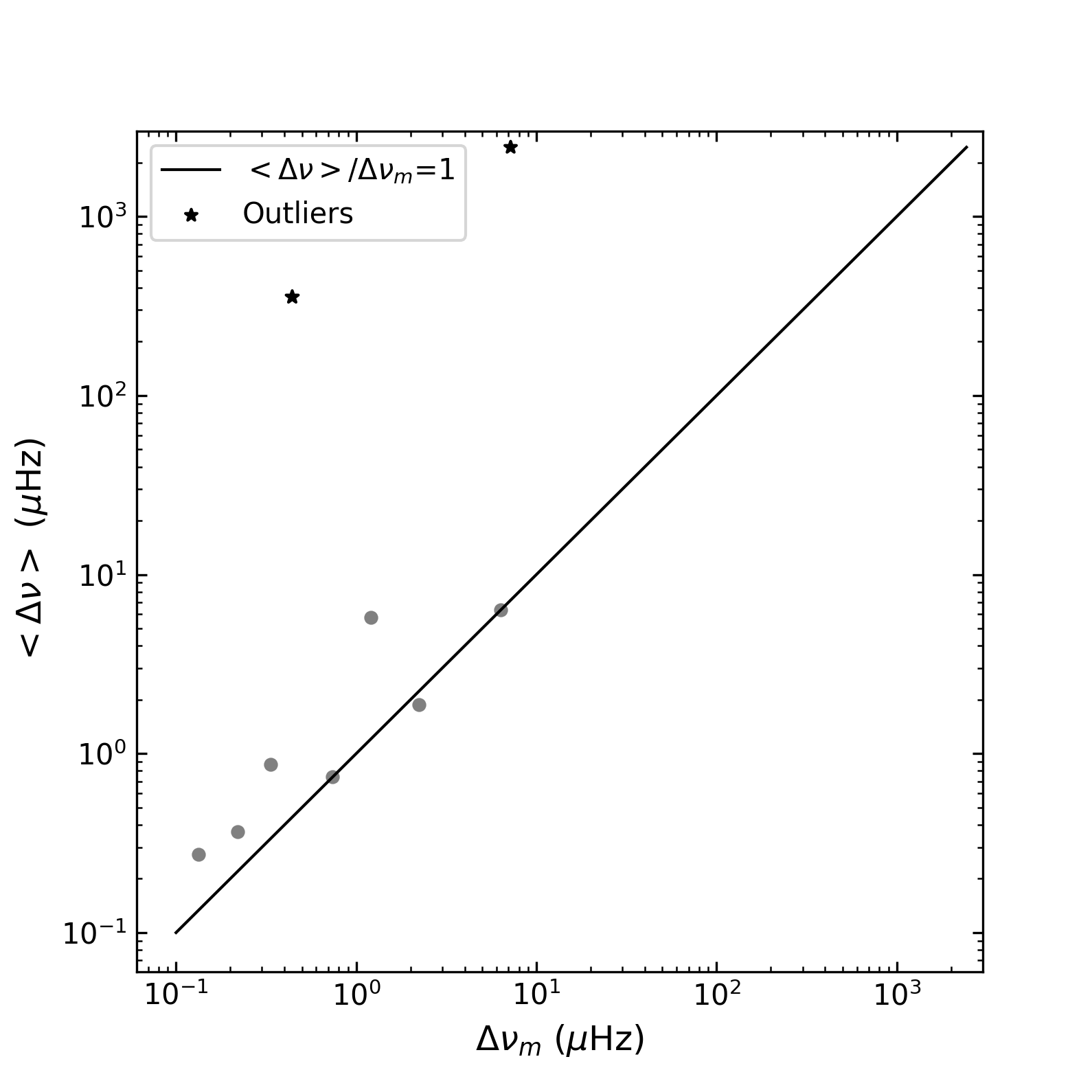} 
 \caption{The average glitch size ⟨$\Delta\nu$⟩ versus the median glitch size $\Delta\nu_{m}$ of nine pulsars are shown as grey points. The marked star points indicate the nine pulsars with a skewed $\Delta\nu$ distribution. The black line follows the relation ⟨$\Delta\nu$⟩=$\Delta\nu_{m}$.}
 \label{9pz}
 \end{figure}

\subsection{Glitch size}
These nine pulsars are all isolated stars with characteristic ages ranging from 0.147 to 3.75 Myr. Large glitches are mostly confined to pulsars with characteristic ages smaller than 10$^5$ yr. Therefore, it is not surprising that all 16 glitches show a small fractional size. Such glitches are difficult to detect for Parkes and Jodrell Bank Observatory and have been missed before \citep{Yu2013}. This could possibly attribute to the high cadence of our observations. The bimodal distribution of the observed fractional glitch sizes was previously reported by several authors \citep{Espinoza2011}. As shown in subplot (c) of Fig.2, the first peak of the distribution lies around 2 $\times$ 10$^{-9}$ and
the second around 10$^{-6}$. As noted by \citet{Espinoza2011}, the left edge of the distribution is significantly limited by observational selection. The actual number of small glitches could be large in the intrinsic distribution. Our observations entirely contribute to the left part of the first peak. The dip at $\Delta\nu/\nu \sim 10^{-7}$ indicates that there may be two mechanisms that can induce a glitch event. It has been proposed that large glitches may result from the sudden transfer of angular momentum from a crustal superfluid to the rest of the star, whereas starquakes caused by the cracking of stellar crust may generate small glitches.  The fractional glitch size depends on both the size of the glitch and the spin frequency of the pulsar. As shown in subplot (a) of Fig.2, $\Delta\nu$ also has a bimodal distribution. However, the peak for large glitches is narrower in $\Delta\nu$ than in $\Delta\nu/\nu$. 

Fig.3 shows the time sequence of fractional glitch sizes for PSR J1833$-$0827, a new frequent glitch pulsar, Crab pulsar, and Vela pulsar. Most glitches in the Vela pulsar (PSR B0833$-$45) are large with similar amplitude, but much smaller glitches were seen occasionally. There are many small glitches and only one large for PSR J1833$-$0827; similar behaviors are also seen in PSRs J1341$-$6220, J1740$-$3015, J0631$+$1036, and J1801$-$2304. Fig.~\ref{9pz} shows the relation between the median $\Delta{\nu}$ ($\Delta{\nu_m}$) and average $\Delta{\nu}$ ($<\Delta{\nu}>$) which indicates the skewness of glitch size for a certain pulsar. Pulsars with symmetrical
glitch size distributions will fall on the straight line, which corresponds to outliers and are considered to be those that lie more than one standard deviation from the straight line and are marked as a star. Outliers shown in Fig.~\ref{9pz} are pulsar that suffers small size glitches but also occasional large events. This is consistent with the results from \cite{Basu2022}.

\subsection{Spin-down rate change}
The change in the rotation frequency during a glitch is often accompanied by a change in the spin-down rate. Negative values of $\Delta\dot{\nu}$ are seen in
the majority of glitches, although the inferred change in the spin-down rate can be either positive or negative. The distribution of $\Delta\dot{\nu}$ is shown in the subplot (b) of Fig.2. The distribution is also bimodal, and our results entirely contribute to the left peak, as was the case for the glitch size distribution. 
The spin-down rate change in $\Delta\dot{\nu}$ is significantly correlate with glitch size \citep{Basu2022}. Therefore, the spin-down rate changes shown in subplot (b) of Fig.2 are also very small.
 
 \subsection{Glitch rate}
 Glitch rate ($R_{g}$) is a useful parameter to get an idea of how active a pulsar is in terms of glitches. This rate might change over the years of observations. The glitch rates are calculated assuming that they are constant in time and should be treated as approximations. The glitch rate ($R_{g}=N/T$) can be defined as where N is the total number of observed glitches, and T is the time interval of observations at the Nanshan and Jodrell Bank Observatory (JBO). The uncertainty on the glitch rates was computed as the square root of N divided by the data span.
 The glitch rate of the nine pulsars derived from the JBO and Nanshan is listed in Table.2. Our data spans are shorter than that of JBO. But the number of glitches we detected is larger than JBO for five pulsars. All the glitch rates obtained from Nanshan are higher than that of JBO. This could possibly attribute to the high cadence of our observations.

\section{Conclusion}
In this paper, we reported glitch events in the timing residuals of 9 pulsars with total data spans of about 114 years. Sixteen new glitches have been identified in these nine pulsars. Glitches have been reported for all nine pulsars before. All 16 glitches show a small fractional size. Some of the 16 glitches may have exponential or linear recovery, but it is challenging for us to make further analyses under the large gap in the data set. The timing accuracy, observational sampling, and intrinsic timing noise may also hamper the detection of post-glitch recoveries with very short time scales. All the glitch rates obtained from Nanshan are higher than that from the Jodrell Bank Observatory. Most known glitches are published by JBO and Parkes. However, such glitches are difficult to detect for Parkes and Jodrell Bank Observatory, and the actual number of small glitches could be large. We also found that PSR J1833$-$0827 is a frequently glitching pulsar with many minor glitches. All the glitch rates obtained from Nanshan are higher than that of JBO. The high glitch rate and small glitch size could possibly result from the high observation cadence.

\begin{acknowledgements}
We would like to thank the XAO pulsar group for discussions and the anonymous referee for helpful suggestions that led to significant improvements in our study.
We are thankful to Prof. XinZhong Er and Prof. Adam Rogers for some useful advice.
The work is supported by the National Natural Science Foundation of China (Grant No.12041304, 11873080, 12033001).
Z.G.W. is supported by the 2021 project Xinjiang Uygur Autonomous Region of China for Tianshan elites, and the National SKA Program of China (Grant No. U1838109, 2020SKA0120100, 12041301).
RY is supported by the Key Laboratory of Xinjiang Uygur Autonomous Region No. 2020D04049, the National SKA Program of China No. 2020SKA0120200, and the 2018 Project of Xinjiang Uygur Autonomous Region of China for Flexibly Fetching in Upscale Talents.
X.Z is supported by CAS ``Light of West China'' Program No. 2018-XBQNXZ-B-025.
The Nanshan 26-m Radio Telescope is partly supported by the Operation, Maintenance and Upgrading Fund for Astronomical Telescopes and Facility Instruments, budgeted from the Ministry of Finance of China (MOF) and administrated by the CAS.

\end{acknowledgements}

\label{lastpage}

\clearpage

\end{document}